\documentclass[%
 reprint,
showpacs,preprintnumbers,
 amsmath,amssymb,
 aps,
]{revtex4-1}

\usepackage{graphicx}
\usepackage{dcolumn}
\usepackage{bm}

\begin{document}


\title{Intrinsic optical conductivity of modified-Dirac fermion systems}
\author{Habib Rostami}
\author{Reza Asgari}
 \email{asgari@ipm.ir}
\affiliation{School of Physics, Institute for Research in Fundamental Sciences (IPM), Tehran 19395-5531, Iran}

\date{\today}

\begin{abstract}
We analytically calculate the intrinsic longitudinal and transverse optical conductivities of electronic
systems which govern by a modified-Dirac fermion model Hamiltonian for materials beyond graphene such as
monolayer MoS$_2$ and ultrathin film of the topological insulator.
We analyze the effect of a topological term in the Hamiltonian on the optical conductivity and transmittance. We show that the optical response enhances in the non-trivial phase of the ultrathin film of the topological insulator and the optical Hall conductivity changes sign at transition from trivial to non-trivial phases which has significant consequences on a circular polarization and optical absorption of the system.
\end{abstract}
\pacs{72.20-i, 78.67.-n, 78.20.-e}
\maketitle
\section{introduction}\label{sect:intro}

Two-dimensional (2D) materials have been one of the most
interesting subjects in condensed matter physics for potential
applications due to the wealth of unusual physical phenomena that
occur when charge, spin and heat transport are confined to a 2D
plane~\cite{Xu}. These materials can be mainly classified in
different classes which can be prepared as a single atom thick
layer namely, layered van der Waals materials, layered ionic
solids, surface growth of monolayer materials, 2D topological
insulator solids and finally 2D artificial systems and they
exhibit novel correlated electronic phenomena ranging from
high-temperature superconductivity, quantum valley or spin Hall
effect to other enormously rich physics phenomena. two-dimensional materials
can be mostly exfoliated into individual thin layers from stacks
of strongly bonded layers with weak interlayer interaction and a
famous example is graphene and hexagonal boron nitride~\cite{Geim}.
The 2D exfoliates versions of transition metal dichalcogenides
exhibit properties that are complementary to and distinct from
those in graphene~\cite{wang12}.

Optical spectroscopy is a broad field and useful to explore the electronic properties of solids. Optical properties can be tuned by varying the
Fermi energy or the electronic band structure of 2D systems.
Recently, developed 2D systems such as gapped
graphene~\cite{Neto09}, thin film of the topological
insulator~\cite{Hasan10,Tibook}, and monolayer of transition
metal dichalcogenides~\cite{wang12} provide the electronic
structures with direct band gap signatures. The optical response of semiconductors with direct band gap is strong
and easy to explore experimentally since photons with energy greater than the energy gap can be absorbed or omitted.
The thin film of the topological insulator, on the other hand, has been
fabricated experimentally by using Sb$_2$Te$_3$
slab~\cite{Zhang13} and has been shown that a direct band gap can be
formed owing to the hybridization of top and bottom surface
states. Furthermore, a non-trivial quantum spin Hall phase has been realized
experimentally which was predicted
previously in this system~\cite{Lu10,Li10,HLi12}.  Although pristine graphene and surface states of the
topological insulator reveal \emph{massless Dirac} fermion physics
, by opening an energy gap they become formed as \emph{ massive Dirac} fermions. The thin film of the topological insulator and
monolayer transition metal dichacogenides can be described by a
modified-Dirac Hamiltonian. A monolayer of the molybdenum
disulfide (ML-MoS$_2$) is a direct band gap
semiconductor~\cite{Mak10}, however its multilayer and bulk show
indirect band gap~\cite{wang12}. This feature causes the
optical response in ML-MoS$_2$ to increases in comparison with its bulk and
multilayer structures~\cite{cao12,cui12,mak12,mak13,wu13}.

One of the main properties of ML-MoS$_2$ is a circular dichroism
aspect responding to a circular polarized light where the left
or right handed polarization of the light couples only to the $K$
or $K'$ valley and it provides an opportunity to induce a valley
polarized excitation which can profoundly be of interest in the application for valleytronics
~\cite{Rycerz07,xiao07,yao08}. Another peculiarity of
ML-MoS$_2$ is the coupled spin-valley in the electronic structure which
is owing to the strong spin-orbit coupling originating from the
existence of a heavy transition metal in the lattice structure and the broken inversion symmetry too.~\cite{xiao12}
These two aspects are captured in a minima massive Dirac-like Hamiltonian
introduced by Xiao et al.~\cite{xiao12} However it has been shown
, based on the tight-binding~\cite{Rostami13,Liu13} and $k.p$
method~\cite{Kormanyos13}, that other terms like an
effective mass asymmetry, a trigonal warping, and a diagonal quadratic term
might be included in the
massive Dirac-like Hamiltonian. The effect of the diagonal quadratic term is
very important, for instance, if the system is exposed by a perpendicular
magnetic field, it will induce a valley degeneracy breaking
term~\cite{Rostami13}. The optical properties of ML-MoS$_2$ have
been evaluated by {\it ab-initio}
calculations~\cite{Carvalho13} and studied theoretically based on the simplified massive Dirac-like model
Hamiltonian~\cite{Li12}, which is by itself valid only near the main absorbtion edge. A part of the model Hamiltonian which describes the dynamic of massive Dirac fermions
are known in graphene committee to have an optical response quite different from that of a standard 2D electron gas. Thus it would be worthwhile to generalize the optical properties of such systems by
using the modified-Dirac fermion model Hamiltonian.

The modified Hamiltonian for ML-MoS$_2$ without trigonal warping effect at $K$ point is very
similar to the \emph{modified-Dirac} equation which has been
studied for an ultrathin film of the topological
insulator (UTF-TI) around $\Gamma$ point.~\cite{Lu10,Shan10} The modified-Dirac
Hamiltonian reveals non-trivial quantum spin hall (QSH) and trivial phases corresponding to the existence and absence of the edge states, respectively. Those phases
have been predicted theoretically~\cite{Lu10,Li10,HLi12,Kim12} and recently
observed by experiment~\cite{Zhang13}. An enhancement of the
optical response of UTF-TI has been obtained in the non-trivial
phase~\cite{peres13} and a band crossing is observed in the
presence of the structure inversion asymmetry induced by
substrate~\cite{Lu13}. Since the modified-Dirac Hamiltonian
incorporates an energy gap and a quadratic term in momentum which both
have topological meaning, it is natural to expect that the topological term of
the Hamiltonian plays an important role in the optical conductivity.
In this paper, we analytically calculate the intrinsic longitudinal and transverse optical conductivities of
the modified-Dirac Hamiltonian as a function of photon energy. This model Hamiltonian covers the main
physical properties of ML-MoS$_2$ and UTF-TI systems in the regime where interband transition plays a main role. We analyze the effect of the topological term
in the Hamiltonian on the optical conductivity and transmittance. Furthermore, we show that the UTF-TI system has a non-trivial
phase and its optical response enhances in addition,
the optical Hall conductivity changes sign at a phase
boundary, when the energy gap is zero. This changing of the sign has a significant consequence on the circular
polarization and the optical absorbtion of the system.

The paper is organized as follows. We introduce the low-energy
model Hamiltonian of ML-MoS$_2$ and UTF-TI systems and then the dynamical conductivity is calculated analytically
by using Kubo formula in Sec.
\ref{sec:theory}. The numerical
results for the optical Hall and longitudinal conductivities and
optical transmittance are reported and we also provide discussions with circular dichroism in both systems in Sec.~\ref{sec:numerical}. A brief summary of results is given in
Sec.~\ref{sec:summary}.

\section{theory and method}\label{sec:theory}

The low-energy properties of the ML-MoS$_2$
and other transition metal dicalcogenide materials can be described by a
modified-Dirac equation~\cite{Rostami13,Kormanyos13,Liu13} and the
Hamiltonian around the $K$ and $K'$ points is given by
\begin{eqnarray}\label{htmos2}
{\cal H}_{\tau s}=\frac{\lambda}{2}\tau s+\frac{\Delta-\lambda\tau s}{2}\sigma_z
+t_0a_0{\bf q}\cdot{\bm \sigma}_\tau
+\frac{\hbar^2|{\bf q}|^2}{4m_0}(\alpha+\beta\sigma_z)\nonumber \\
\end{eqnarray}
where the Pauli matrices stand for a pseudospin which indicates the conduction and valence band degrees of freedom, $\tau=\pm$ denotes the two independent valleys in the first Brillouin zone, ${\bf q}=(q_x,q_y)$ and ${\bm \sigma}_{\tau}=(\tau \sigma_x,\sigma_y)$. The numerical values of the parameters will be given in the Sec.~\ref{sec:numerical}.

The UTF-TI system, on the other hand, can be described by a modified Dirac Hamiltonian around the $\Gamma$ point with two independent hyperbola (isospin) degree of freedoms~\cite{Lu10,Shan10} and thus the Hamiltonian reads
\begin{eqnarray}\label{hti}
{\cal H}_{\tau}=\epsilon_0+\tau\frac{\Delta}{2}\sigma_z+t_0a_0{\bf q}\cdot{\bm \sigma}+\frac{\hbar^2|{\bf q}|^2}{4m_0}(\alpha+\tau\beta\sigma_z)
\end{eqnarray}
Note that the Pauli matrices in this Hamiltonian stand for the real spin where spin is rotated by operator $U={\rm diag}[1,i]$ which results in $U^\dagger\sigma_xU=-\sigma_y$ and $U^\dagger\sigma_yU=\sigma_x$ and the isospin index of $\tau=\pm$ indicates two independent solutions of UTF-TI which are degenerated in the absence of the structure inversion asymmetry and can be assumed as an internal isospin (spin, valley, or sublattice) degree of freedom. Two mentioned models, Eqs.~(\ref{htmos2}) and (\ref{hti}), are similar to some extent and describe similar physical properties.

Generally, the Hamiltonian around the $\Gamma(\tau=+)$ and $K(\tau=+$) points for UTF-TI and monolayer MoS$_2$ systems, respectively can be re-written as
\begin{eqnarray}\label{hamiltonain}
H=\begin{pmatrix}a_1+b(\alpha+\beta)q^2&&c q^\ast\\c q && a_2+b(\alpha-\beta)q^2\end{pmatrix}
\end{eqnarray}
where $a_1=\Delta/2+\epsilon_0,~a_2=-\Delta/2+\epsilon_0$ for UTF-TI and $a_1=\Delta/2,~a_2=-\Delta/2+\lambda s$ for ML-MoS$_2$. Note that $b=\hbar^2/4m_0a_0^2,~c=t_0$, and we set $a_0q\rightarrow q$. The eigenvalue and eigenvector of the Hamiltonian, Eq.~(\ref{hamiltonain}) can be obtained as
\begin{eqnarray}
&&|\psi_{c,v}\rangle=\frac{1}{D_{c,v}}\begin{pmatrix}-c q^\ast\\h_{c,v}\end{pmatrix}\nonumber\\&&
h_{c,v}=d\mp\sqrt{d^2+c^2 q^2}~~~,~~~
d=\frac{a_1-a_2}{2}+b\beta q^2\nonumber\\&&
D_{c,v}=\sqrt{c^2 q^2+h^2_{c,v}}\nonumber\\&&
\varepsilon_{c,v}=a_1+b(\alpha+\beta)q^2-h_{c,v}
\end{eqnarray}
and velocity operators along the $x$ and $y$ directions are
\begin{eqnarray}
\hbar v_x=\frac{\partial H}{\partial q_x}=c\sigma_x+2b\alpha q_x+2b\beta q_x\sigma_z\nonumber\\
\hbar v_y=\frac{\partial H}{\partial q_y}=c\sigma_y+2b\alpha q_y+2b\beta q_y\sigma_z
\end{eqnarray}
The intrinsic optical conductivity can be calculated by using the Kubo formula~\cite{Stauber08,Tse11,Louie06} in a clean sample and it is given by\begin{widetext}
\begin{eqnarray}
&&\sigma_{xy}(\omega)=-i\frac{e^2}{2\pi h}\int{d^2q\frac{f(\varepsilon_c)-f(\varepsilon_v)}{\varepsilon_c-\varepsilon_v}
\{\frac{\langle\psi_c|\hbar v_x
|\psi_v\rangle\langle\psi_v|\hbar v_y |\psi_c\rangle}{\hbar\omega+\varepsilon_c-\varepsilon_v+i0^+}
+\frac{\langle\psi_v|\hbar v_x |\psi_c\rangle\langle\psi_c|\hbar v_y |\psi_v\rangle}{\hbar\omega+\varepsilon_v-\varepsilon_c+i0^+}\}}\nonumber\\&&
\sigma_{xx}(\omega)=-i\frac{e^2}{2\pi h}\int{d^2q\frac{f(\varepsilon_c)-f(\varepsilon_v)}{\varepsilon_c-\varepsilon_v}
\{\frac{\langle\psi_c|\hbar v_x
|\psi_v\rangle\langle\psi_v|\hbar v_x |\psi_c\rangle}{\hbar\omega+\varepsilon_c-\varepsilon_v+i0^+}
+\frac{\langle\psi_v|\hbar v_x |\psi_c\rangle\langle\psi_c|\hbar v_x |\psi_v\rangle}{\hbar\omega+\varepsilon_v-\varepsilon_c+i0^+}\}}
\end{eqnarray}
\end{widetext}
where $f(\omega)$ is the Fermi distribution function.
We include only the interband transitions and the contribution of the intraband transitions, which leads to the fact that the Drude-like term, is no longer relevant in this study since the momentum relaxation time is assumed to be infinite. This approximation is valid at low-temperature and a clean sample where defect, impurity, and phonon scattering mechanisms are ignorable. We also do not consider the bound state of exciton in the systems. After straightforward calculations (details can be found in Appendix A), the real and imaginary parts of diagonal and off-diagonal
components of the conductivity tensor at $\tau=+$ are given by
\begin{widetext}
\begin{eqnarray}\label{general}
&&\sigma^{\Re}_{xy}(\omega)=\frac{2e^2}{h}\int{qdq (f(\varepsilon_c)-f(\varepsilon_v))\times\{\frac{c^2}{\sqrt{d^2+c^2q^2}}(d-2b\beta q^2)\}
\{\mathbb{P}\frac{-1}{(\hbar\omega)^2-(\varepsilon_c-\varepsilon_v)^2}\}}\nonumber\\&&
\sigma^{\Im}_{xy}(\omega)=\frac{\pi e^2}{h}\int{qdq \frac{f(\varepsilon_c)-f(\varepsilon_v)}{\varepsilon_c-\varepsilon_v}
\times\{\frac{c^2}{\sqrt{d^2+c^2q^2}}(d-2b\beta q^2)\}
\{\delta(\hbar\omega+\varepsilon_v-\varepsilon_c)-\delta(\hbar\omega+\varepsilon_c-\varepsilon_v)\}}\nonumber\\&&
\sigma^{\Im}_{xx}(\omega)=-\frac{2e^2}{h}\hbar\omega\int{qdq \frac{f(\varepsilon_c)-f(\varepsilon_v)}{\varepsilon_c-\varepsilon_v}\times\{ c^2-\frac{c^2q^2}{d^2+c^2q^2}[\frac{c^2}{2}+b\beta(a_1-a_2)]\}
\{\mathbb{P}\frac{-1}{(\hbar\omega)^2-(\varepsilon_c-\varepsilon_v)^2}\}}\nonumber\\&&
\sigma^{\Re}_{xx}(\omega)=-\frac{\pi e^2}{h}\int{qdq \frac{f(\varepsilon_c)-f(\varepsilon_v)}{\varepsilon_c-\varepsilon_v}
\times\{ c^2-\frac{c^2q^2}{d^2+c^2q^2}[\frac{c^2}{2}+b\beta(a_1-a_2)]\}
\{\delta(\hbar\omega+\varepsilon_v-\varepsilon_c)+\delta(\hbar\omega+\varepsilon_c-\varepsilon_v)\}}
\end{eqnarray}
\end{widetext}
where $\Re$ and $\Im$ refer to the real and imaginary parts of $\sigma$ and $\mathbb{P}$ denotes the principle value.
It is worthwhile mentioning that the conductivity for ML-MoS$_2$ for $\tau=-$ can be found by implementing $p_x\rightarrow -p_x$ and $\lambda\rightarrow
-\lambda$. Using these transformations, the velocity matrix
elements around the $K'$ point can be calculated by taking
the complex conjugation of the corresponding results for the $\tau=+$ case. Furthermore, for the UTF-TI case system, we must replace $\Delta$ and $\beta$ by their opposite signs which lead to the same results in comparison with the ML-MoS$_2$ case around $K'$ point. More details in this regard are given in Appendix A.

\subsection{Optical conductivity of ML-MoS$_2$}
Having obtained the general expressions of the conductivity for the modified-Dirac fermion systems, the conductivity of two examples namely the ML-MoS$_2$ and UTF-TI could be obtained. Here, we would like to focus on the ML-MoS$_2$ case and explore its optical properties, although all results can be generalized to the UTF-TI system as well. Therefore, the optical conductivity for each spin and valley components of ML-MoS$_2$ can be obtained by using appropriate substitution in Eq.~(\ref{general}) and results are written as
\begin{widetext}
\begin{eqnarray}\label{conductivity}
&&\sigma^{\Re,\tau s}_{xy}(\omega)=\frac{2e^2}{h}\mathbb{P}\int{dq (f(\varepsilon_c)-f(\varepsilon_v))\times\{\frac{\tau(\Delta'_{\tau s}q-\beta'q^3)}{\sqrt{(\Delta'_{\tau s}+\beta'q^2)^2+q^2}[4((\Delta'_{\tau s}+\beta'q^2)^2+q^2)-(\hbar\omega/t_0)^2]}\}}
\nonumber\\ \nonumber\\&&
\sigma^{\Im,\tau s}_{xy}(\omega)=\frac{\pi e^2}{2h}\int{dq (f(\varepsilon_c)-f(\varepsilon_v))
\times\{\frac{\tau(\Delta'_{\tau s} q-\beta'
q^3)}{(\Delta'_{\tau s}+\beta'q^2)^2+q^2}\}
\delta(\hbar\omega/t_0-2\sqrt{(\Delta'_{\tau s}+\beta'q^2)^2+q^2})}\nonumber\\ \nonumber\\&&
\sigma^{\Im,\tau s}_{xx}(\omega)=-\frac{2e^2}{h}\hbar\omega \mathbb{P}\int{dq (f(\varepsilon_c)-f(\varepsilon_v))\times\{ \frac{q}{\sqrt{(\Delta'_{\tau s}+\beta'q^2)^2+q^2}[4((\Delta'_{\tau s}+\beta'q^2)^2+q^2)-(\hbar\omega/t_0)^2]}}\nonumber\\&&
~~~~~~~~~~~~~~~~~~~~~~~~~~~~~~~~~~~~~~~~~~~~~~~~~~~~~-{\frac{q^3[\frac{1}{2}+2\beta'\Delta'_{\tau s}]}{((\Delta'_{\tau s}+\beta'q^2)^2+q^2)^{3/2}
[4((\Delta'_{\tau s}+\beta'q^2)^2+q^2)-(\hbar\omega/t_0)^2]}\}}\nonumber\\ \nonumber\\&&
\sigma^{\Re,\tau s}_{xx}(\omega)=-\frac{\pi e^2}{2h}\int{dq (f(\varepsilon_c)-f(\varepsilon_v))\times\{ \frac{q}{\sqrt{(\Delta'_{\tau s}+\beta'q^2)^2+q^2}}}
-{\frac{q^3[\frac{1}{2}+2\beta'\Delta'_{\tau s}]}{((\Delta'_{\tau s}+\beta'q^2)^2+q^2)^{3/2}}\}
\delta(\hbar\omega/t_0-2\sqrt{(\Delta'_{\tau s}+\beta'q^2)^2+q^2})}\nonumber\\
\end{eqnarray}
\end{widetext}
where $\Delta'_{\tau s}=(\Delta-\lambda\tau s)/2t_0$,  $\alpha'=b\alpha/t_0$, $\beta'=b\beta/t_0$, $\sigma^{\tau,s}_{xy}=\sigma^{\Re,\tau s}_{xy}+i\sigma^{\Im,\tau s}_{xy}$, and $\sigma^{\tau,s}_{xx}=\sigma^{\Re,\tau s}_{xx}+i\sigma^{\Im,\tau s}_{xx}$.

Note that in the case of UTF-TI, there is no extra spin index of $s$ as a degree of freedom and $\Delta'_{\tau s}$ might be replaced by $\Delta'=\Delta/2t_0$, consequently we have $\sigma^{\tau}_{xy}$ and $\sigma^{\tau}_{xx}$ rather than $\sigma^{\tau s}_{xy}(\omega)$ and $\sigma^{\tau s}_{xx}(\omega)$. To be more precise, $\lambda\tau s$, which is located out of the radical in Eq.~(\ref{conductivity}), might be replaced by $\epsilon_0$ in the $\epsilon_{c,v}$ to achieve desirable results corresponding to the UTF-TI. It is clear that the dynamical charge Hall conductivity vanishes in both the UTF-TI and ML-MoS$_2$ systems due to the presence of the time reversal symmetry. For the MoS$_2$ case, the spin and valley \emph{transverse} ac-conductivity are given by
\begin{eqnarray}
&&\sigma^s_{xy}=\frac{\hbar}{2e}\sum_{\tau}[\sigma^{\tau,\uparrow}_{xy}-\sigma^{\tau,\downarrow}_{xy}]\nonumber\\&&
\sigma^v_{xy}=\frac{1}{e}\sum_{s}[\sigma^{K,s}_{xy}-\sigma^{K',s}_{xy}]
\end{eqnarray}
and for the \emph{longitudinal} ac-conductivity case, an electric field can only induce a charge current and corresponding conductivity is given as
\begin{eqnarray}\label{sigmaxx}
\sigma_{xx}=\sum_{\tau}[\sigma^{\tau,\uparrow}_{xx}+\sigma^{\tau,\downarrow}_{xx}]
\end{eqnarray}
Moreover, the longitudinal conductivity is the same as expression given by Eq.~(\ref{sigmaxx}) for the UTF-TI case however, the Hall conductivity is  slightly changed.
Owing to the coupling between the isospin and the spin indexes, the hyperbola Hall conductivity is a spin Hall conductivity~\cite{Lu10,Zhang13} and it is thus given by
\begin{eqnarray}
\sigma^{hyp}_{xy}=\frac{1}{e}[\sigma^{\Gamma^+}_{xy}-\sigma^{\Gamma^-}_{xy}]
\end{eqnarray}

\subsection{Intrinsic dc-conductivity}
To find the static conductivity in a clean sample, we set $\omega=0$ and thus the interband longitudinal
conductivity vanishes. Consequently, we calculate only
the transverse conductivity in this case. At zero temperature, the Fermi distribution function is
given by a step function, {\it i. e.}
$f(\varepsilon_{c,v})=\Theta(\varepsilon_{\rm
F}-\varepsilon_{c,v})$. We derive the optical conductivities for the case of
ML-MoS$_2$ and results corresponding to the UTF-TI can be
deduced from those after appropriate substitutions. Most of the
interesting transport properties of ML-MoS$_2$ originates from its
spin splitting band structure for the hole doped case. Therefore,
for the later case, when the upper spin-split band contributes to
the Fermi level state, the dc-conductivity is given by
\begin{eqnarray}
 &&\sigma^{K \uparrow}_{xy}=-\sigma^{K'\downarrow}_{xy}=-\frac{e^2}{2h}\int^{q_{c}}_{q_{\rm F}}{\frac{(\Delta'_{K\uparrow}q-\beta'q^3) dq}{((\Delta'_{K\uparrow}+\beta'q^2)^2+q^2)^{\frac{3}{2}}}}\nonumber\\&&
=-\frac{e^2}{2h}{\cal C}^{K\uparrow}+\frac{e^2}{2h}\frac{2\mu+2b(\alpha-\beta)q^2_{\rm F}}{\Delta-\lambda+2\mu+2b\alpha q^2_{\rm F}}
 \end{eqnarray}
and for the spin-down component we thus have
\begin{eqnarray}
\sigma^{K \downarrow}_{xy}&=&-\sigma^{K'\uparrow}_{xy}=-\frac{e^2}{2h}\int^{q_{c}}_{0}{\frac{(\Delta'_{K\downarrow}q-\beta'q^3) dq}{((\Delta'_{K\downarrow}+\beta'q^2)^2+q^2)^{\frac{3}{2}}}}\nonumber\\
&=&-\frac{e^2}{2h}{\cal C}^{K\downarrow}
\end{eqnarray}
where $q_c$ is the ultra violate cutoff and $\mu/t_0=\sqrt{(\Delta'_{K\uparrow}+\beta'q^2_{\rm
F})^2+q^2_{\rm F}}-\Delta'_{K\uparrow}-\alpha'q^2_{\rm F}$ stands
for the chemical potential and it is easy to show that ${\cal
C}^{Ks}={\rm sgn}(\Delta-\lambda s)-{\rm sgn}(\beta)$ at large
cutoff values. In a precise definition, ${\cal C}^{Ks}$ terms are the Chern
numbers for each spin and valley degrees of freedom and the total
Chern number is zero owing to the time reversal symmetry.
Intriguingly, the quadratic term in Eq.~(3), $\beta$, leads to a new topological
characteristic. When $\beta\Delta>0$, with $\Delta>\lambda$,
system has a trivial phase with no edge mode closing the energy
gap however for the case that $\beta\Delta<0$, the topological
phase of the system is a non-trivial with edge modes closing the energy gap. In the case of the ML-MoS$_2$, the tight binding
model~\cite{Rostami13,Gomez13} predicts the trivial
phase ($\beta>0)$ with ${\cal C}^{Ks}=0$. However, a non-trivial phase is expected by Refs.~[\onlinecite{Kormanyos13}, \onlinecite{Liu13}] (where $\beta<0$) which leads to ${\cal C}^{Ks}=2$. In other words, the
term proportional to $\beta$ has a topological meaning in Z$_2$
symmetry invariant like the UTF-TI system~\cite{Lu10} and the sign of $\beta$ plays important role.

The transverse intrinsic dc-conductivity for the hole doped ML-MoS$_2$ case, is given by
\begin{eqnarray}\label{sh1}
\sigma^s_{xy}&=&\frac{\hbar}{e}[\sigma^{K\uparrow}_{xy}-\sigma^{K\downarrow}_{xy}]=\frac{e}{2\pi}\frac{\mu+b(\alpha-\beta)q^2_{\rm F}}{\Delta-\lambda+2\mu+2b\alpha q^2_{\rm F}}\nonumber\\
\sigma^v_{xy}&=&\frac{2}{e}[\sigma^{K\uparrow}_{xy}+\sigma^{K\downarrow}_{xy}]=-\frac{e}{h}{\cal C}^{K}+\frac{2}{\hbar}\sigma^s_{xy}
\end{eqnarray}
where, at large cutoff, ${\cal C}^{K}=[{\rm
sign}(\Delta-\lambda)+{\rm sign}(\Delta+\lambda)]/2-{\rm
sign}(\beta)$ stands for the valley Chern number and it equals to
zero or $2$ corresponding to the non-trivial or trivial band
structure, respectively. In the case of the UTF-TI, the isospin Hall conductivity is
\begin{eqnarray}
\sigma^{hyp}_{xy}=-\frac{e}{h}{\cal C}^{\Gamma}+\frac{2e}{h}\frac{\mu+b(\alpha-\beta)q^2_{\rm F}}{\Delta+2\mu+2b\alpha q^2_{\rm F}}
\end{eqnarray}
where $\mu/t_0=\sqrt{(\Delta'+\beta'q^2_{\rm F})^2+q^2_{\rm
F}}-\Delta'-\epsilon_0-\alpha'q^2_{\rm F}$ and  ${\cal
C}^{\Gamma}={\rm sgn}(\Delta)-{\rm sgn}(\beta)$ at large cutoff.
This result is consistent with that result obtained by Lu {\it et
al.}~\cite{Lu10}. It should be noted that in the absence
of the diagonal quadratic term, the non-zero valley Chern number
at zero doping predicts a valley Hall conductivity, which is
proportional to ${\rm sign}(\Delta)$. Therefore, the exitance of edge states
, which can carry the valley current, is anticipated. However, Z$_2$ symmetry prevents the
edge modes from existing. Since the Z$_2$ topological invariant is zero
when the gap is caused only the inversion symmetry
breaking~\cite{yao09}, thus the topology of the band structure is
trivial and there are no edge states to carry the valley
current when the chemical potential is located inside the energy gap. Therefore, we can ignore the valley Chern number in
$\sigma^v_{xy}$ and thus the results are consistent with those results
reported by Xiao {\it el al.}~\cite{xiao12} at a low doping rate where
$\mu\ll\Delta-\lambda$.

\subsection{Intrinsic dynamical conductivity}

In this section, we analytically calculate the dynamical
conductivity of the modified-Dirac Hamiltonian which results in
the trivial and non-trivial phases. Using the two-band Hamiltonian,
including the quadratic term in momentum, the optical Hall conductivity for
each spin and valley components are given by
\begin{eqnarray}\label{sxy}
\sigma^{\Re,\tau s}_{xy}(\omega)&=&\tau\frac{e^2}{h}[G_{\tau s}(\omega,q_{\rm F})-G_{\tau s}(\omega,q_c)]\nonumber\\
\sigma^{\Im,\tau s}_{xy}(\omega)&=&\tau \frac{\pi e^2}{2h}\frac{\Delta'_{\tau s}-\beta' q_{0,\tau s}^2}{\hbar\omega'n(\omega')}\nonumber\\ &\times& [\Theta(2\varepsilon'_{\rm F}-\lambda'\tau s-2\alpha' q_{0,\tau s}^2-\hbar\omega')-(\omega'\rightarrow -\omega')]\nonumber\\ &\times&\Theta(n(\omega')-(1+2\beta'\Delta'_{\tau s}))
\end{eqnarray}
where $\Re$ and $\Im$ indicate to the real and imaginary parts, respectively and $G_{\tau s}(\omega,q)$ reads as below (details are given in
Appendix B)
\begin{eqnarray}\label{GG}
&&G_{\tau s}(\omega,q)=\frac{\Delta'_{\tau s}}{\hbar\omega' n(\omega')}\ln|\frac{\hbar\omega' \frac{m(q)}{n(\omega')}-2\sqrt{(\Delta'_{\tau s}+\beta' q^2)^2+q^2}}{\hbar\omega' \frac{m(q)}{n(\omega')}+2\sqrt{(\Delta'_{\tau s}+\beta'q^2)^2+q^2}}|\nonumber\\
&&+\frac{1}{4\beta'\hbar\omega' n(\omega')}\ln|\frac{\hbar\omega' \frac{m(q)}{n(\omega')}-2\sqrt{(\Delta'_{\tau s}+\beta'q^2)^2+q^2}}{\hbar\omega'\frac{m(q)}{n(\omega')}+2\sqrt{(\Delta'_{\tau s}+\beta' q^2)^2+q^2}}|\nonumber\\
&&-\frac{1}{4\beta'\hbar\omega'}\ln|\frac{\hbar\omega'-2\sqrt{(\Delta'_{\tau s}+\beta' q^2)^2+q^2}}{\hbar\omega'+2\sqrt{(\Delta'_{\tau s}+\beta' q^2)^2+q^2}}|
\end{eqnarray}
where $m(q)=1+2\beta'\Delta'_{\tau s}+2\beta'^2q^2$,
$n(\omega')=\sqrt{1+4\beta'\Delta'_{\tau
s}+\beta'^2(\hbar\omega')^2}$, $\hbar\omega'=\hbar\omega/t_0$,
$\varepsilon'_{\rm F}=\varepsilon_{\rm F}/t_0$ and
$\lambda'=\lambda/t_0$. The value of $q_{0,\tau s}$ can be
evaluated from $m(q_{0,\tau s})=n(\omega')$. Note that $q_c$, the ultra violate cutoff, is assumed to be equal to $1/a_0$.
Some special attentions might be taken for the situation in which there is no intersection between the Fermi energy and the
band energy, for instance in a low doping hole case of the ML-MoS$_2$ in which the Fermi
energy lies in the spin-orbit splitting interval. In this case,
the Fermi wave vector ($q_{\rm F}$, which has no contribution to the Fermi level) vanishes.

The quadratic terms can also affect profoundly on the longitudinal
dynamical conductivity which plays main role in
the optical response when the time reversal symmetry is preserved.
In this case, one can find
\begin{eqnarray}\label{sxx}
\sigma^{\Re,\tau s}_{xx}(\omega)&=&-\frac{\pi e^2}{4h}\frac{1}{n(\omega')}(1-\frac{1+4\beta'\Delta'_{\tau s}}{2}(\frac{2q_{0,\tau s}}{\hbar\omega'})^2) \nonumber\\&\times& [\Theta(2\varepsilon'_{\rm F}-\lambda'\tau s-2\alpha' q_{0,\tau s}^2-\hbar\omega')-(\omega'\rightarrow -\omega')]\nonumber\\ &\times&\Theta(n(\omega')-(1+2\beta'\Delta'_{\tau s}))\nonumber\\
\sigma^{\Im,\tau s}_{xx}(\omega)&=&-\frac{e^2}{h}[H_{\tau s}(\omega,q_{\rm F})-H_{\tau s}(\omega,q_c)]
\end{eqnarray}
where $H_{\tau s}(\omega,q)$ is given by (details are given in
Appendix B)
\begin{eqnarray}\label{HH}
H_{\tau s}(\omega,q)&=&\frac{(1+2\beta'\Delta'_{\tau s})m(q)-(1+4\beta'\Delta'_{\tau s})}{2\beta'^2\hbar\omega'\sqrt{(\Delta'_{\tau s}+\beta'q^2)^2+q^2}}\nonumber\\
&+&\frac{1+4\beta'\Delta'_{\tau s}}{2\beta'^2(\hbar\omega')^2}\ln|\frac{\frac{\hbar\omega'}{2}-\sqrt{(\Delta'_{\tau s}+\beta'q^2)^2+q^2}}{\frac{\hbar\omega'}{2}+\sqrt{(\Delta'_{\tau s}+\beta'q^2)^2+q^2}}|\nonumber\\
&+&\frac{(1+2\beta'\Delta'_{\tau s})(1+4\beta'\Delta'_{\tau s})+\beta'^2(\hbar\omega')^2}{2\beta'^2(\hbar\omega')^2n(\omega')}\nonumber\\
&\times&\ln|\frac{\frac{\hbar\omega'}{2}\frac{m(q)}{n(\omega')}-\sqrt{(\Delta'_{\tau  s}+\beta'q^2)^2+q^2}}{\frac{\hbar\omega'}{2}\frac{m(q)}{n(\omega')}+\sqrt{(\Delta'_{\tau s}+\beta'q^2)^2+q^2}}|
\end{eqnarray}
It is worthwhile mentioning that the $G$ and $H$
functions do not depend on the $\alpha$ term given in Eq.~(3). For
$\beta=0$ in Eq.~(3), we have $m(q)/n(\omega')\rightarrow 1$,
$1/n(\omega')\rightarrow1-2\beta'\Delta'_{\tau s}$, therefore
$G_{\tau s}(\omega,q)$ reduces to $g_{\tau s}(\omega,q)$ and in a
similar way, $H_{\tau s}$ reduces to
$h_{\tau s}$. Here $g_{\tau s}$ and $h_{\tau s}$ read as below
\begin{eqnarray}\label{gh}
&&g_{\tau s}(\omega,q)=\frac{\Delta-\lambda\tau s}{4\hbar\omega}\ln|\frac{\hbar\omega-\sqrt{(\Delta-\lambda\tau s)^2+4t^2_0q^2}}{\hbar\omega+\sqrt{(\Delta-\lambda\tau s)^2+4t^2_0q^2}}|\nonumber\\
&&h_{\tau s}(\omega,q)=\frac{\Delta-\lambda\tau s}{2\hbar\omega}\frac{\Delta-\lambda\tau s}{\sqrt{(\Delta-\lambda\tau s)^2+4t^2_0q^2}}\nonumber\\&&+\frac{1}{4}[1+(\frac{\Delta-\lambda\tau s}{\hbar\omega})^2]\ln|\frac{\hbar\omega-\sqrt{(\Delta-\lambda\tau s)^2+4t^2_0q^2}}{\hbar\omega+\sqrt{(\Delta-\lambda\tau s)^2+4t^2_0q^2}}|\nonumber\\
\end{eqnarray}
Using Eqs.~(\ref{conductivity}) and (\ref{gh}), the conductivity simplifies when $\beta=0$ and the results are
\begin{eqnarray}\label{sigmaxy1}
\sigma^{\Re,\tau s}_{xy}(\omega)&=&\tau\frac{e^2}{h}[g_{\tau s}(\omega,q_{\rm F})-g_{\tau s}(\omega,q_c)]\nonumber\\
\sigma^{\Im,\tau s}_{xy}(\omega)&=&\tau \frac{\pi e^2}{4h}\frac{\Delta-\lambda\tau s}{\hbar\omega}[\Theta(2\varepsilon_{\rm F}-\lambda\tau s-\hbar\omega)-(\omega\rightarrow -\omega)]\nonumber\\ &\times& \Theta(\hbar\omega-(\Delta-\lambda\tau s))
\end{eqnarray}
The longitudinal conductivity for the case of $\beta=0$ is given by the following relations
for the electron doped case
\begin{eqnarray}\label{sigmaxx1}
\sigma^{\Re,\tau s}_{xx}(\omega)&=&-\frac{\pi e^2}{8h}(1+(\frac{\Delta-\lambda\tau s}{\hbar\omega})^2)\Theta(\hbar\omega-(\Delta-\lambda\tau s))\nonumber\\ &\times&[\Theta(2\varepsilon_{\rm F}-\lambda\tau s-\hbar\omega)-(\omega\rightarrow -\omega)]\nonumber\\
\sigma^{\Im,\tau s}_{xx}(\omega)&=&-\frac{e^2}{h}[h_{\tau s}(\omega,q_{\rm F})-h_{\tau s}(\omega,q_c)]
\end{eqnarray}
These relations are consistent with those results reported
in Ref.~[\onlinecite{Li12}]. Furthermore, dropping the $\lambda$ term gives rise to the
optical conductivity of gapped graphene and the result is in good
agreement with the universal conductivity of
graphene~\cite{Zigler07} for $\Delta=\lambda=\alpha=\beta=0$.

\section{numerical results}\label{sec:numerical}

In most numerical results, we use $set_0: \lambda=0.08eV,~\Delta=1.9eV, t_0=1.68eV,~\alpha=m_0/m_{+}=0.43,~\beta=m_0/m_{-}-4m_0v^2/(\Delta-\lambda)=2.21$ where $m_{\pm}=m_e m_h/(m_h \pm m_e)$ and $v=t_0 a_0/\hbar$. These values have been obtained in
Ref.~[\onlinecite{Rostami13}]. Moreover, for the sake of completeness, we
introduce two other sets of the parameters as $ t_0=1.51eV,
~\beta=1.77$ and another set $t_0=2.02eV,~\beta=0$ corresponding to the
same effective masses ($\alpha=0$ for $m_e=-m_h=0.5 m_0$) for electron and hole bands.
These parameters are calculated by using the procedure reported in
Ref.~[\onlinecite{Rostami13}]. The later comparison helps us to perceive
the validity of the effective mass approximation for the ML-MoS$_2$ system and for this purpose, we
assume the same effective masses for electron and hole bands to compare the spin Hall conductivity resulted from the Dirac-like and modified-Dirac Hamiltonians. Notice that all energies are measured from the center of the energy gap.

\begin{figure}
\includegraphics[width=1\linewidth]{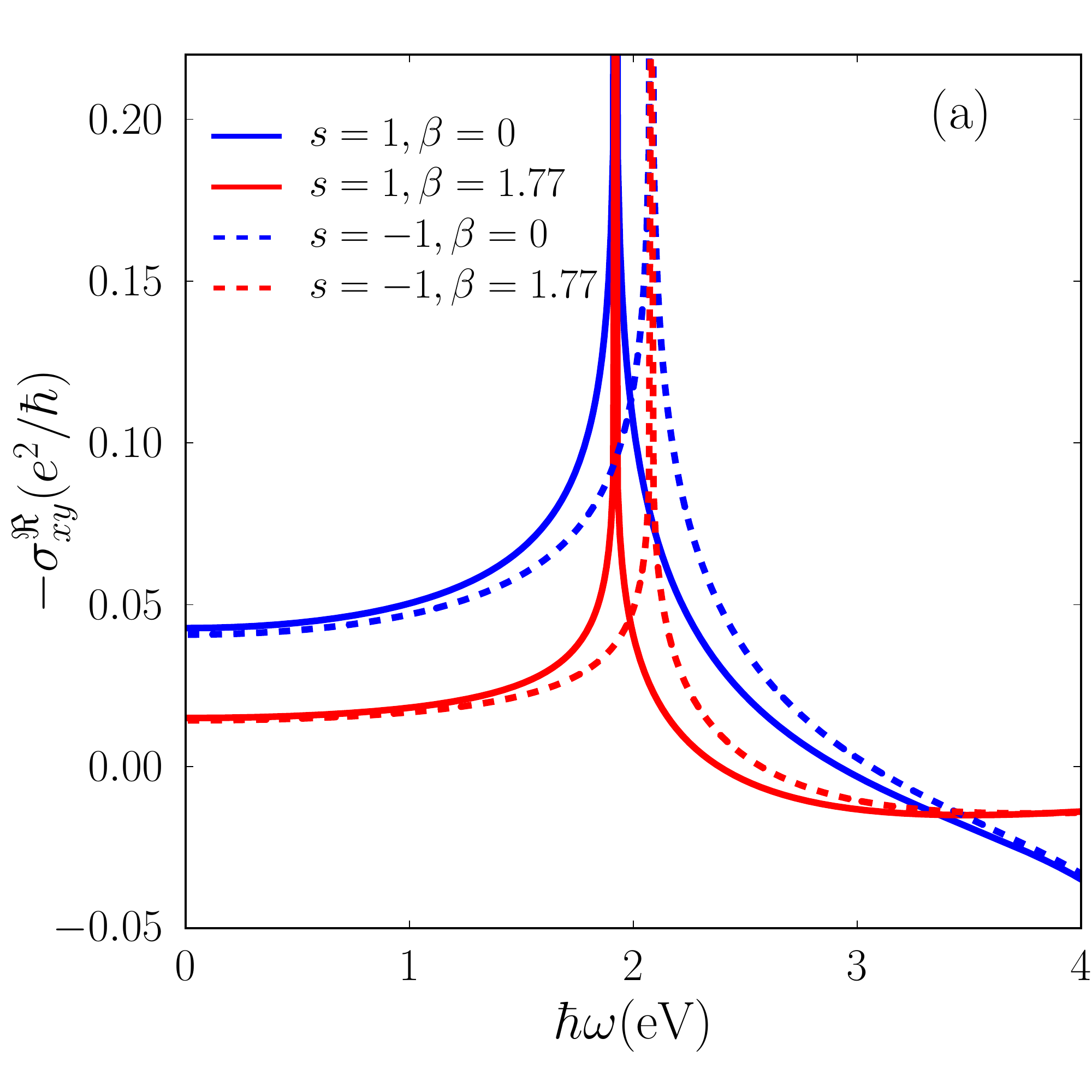}
\includegraphics[width=1\linewidth]{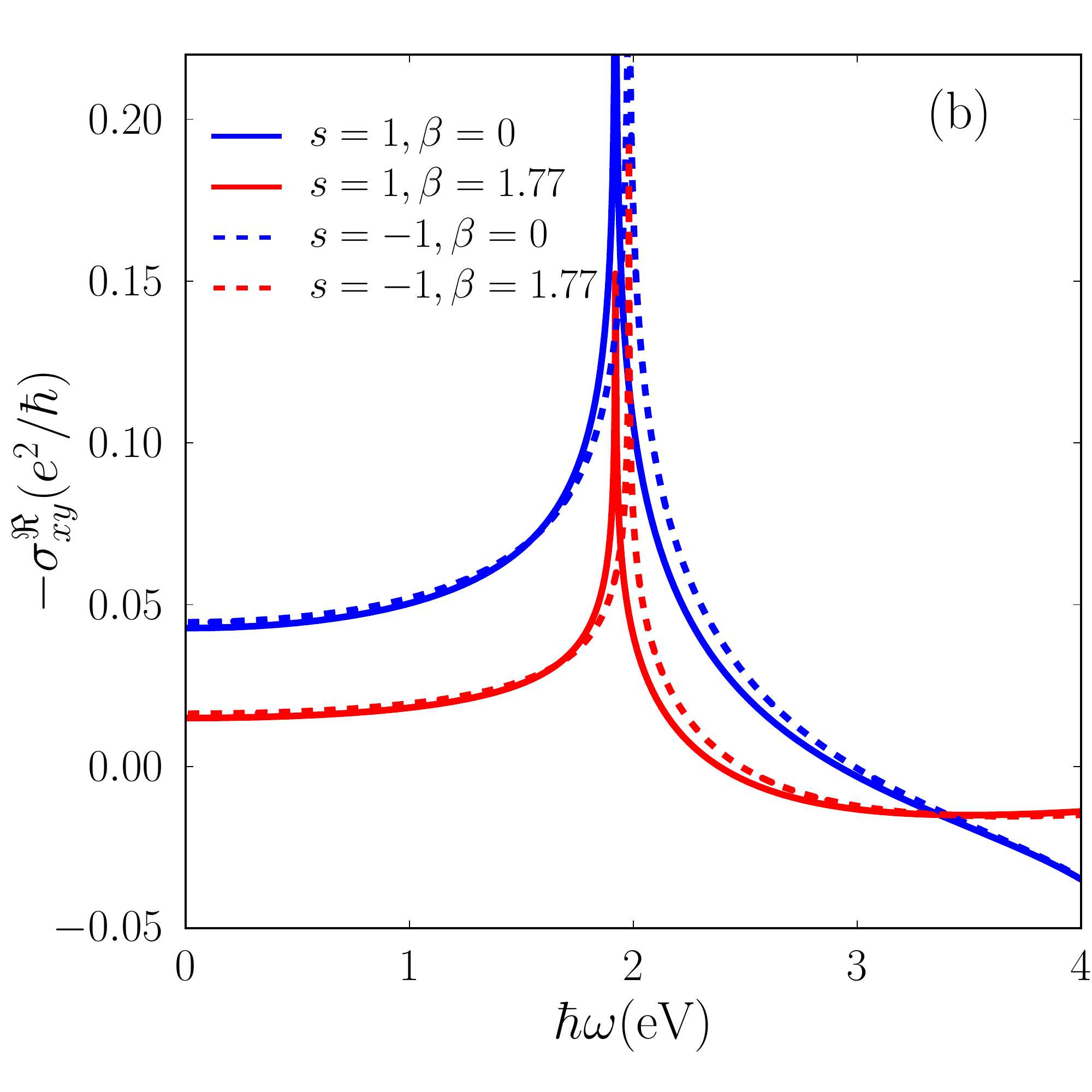}
\caption{(Color online). Real part of the Hall conductivity (in units of $e^2/\hbar$)
for (a) electron with $\varepsilon_{\rm F}=1eV$ and (b) hole with $\varepsilon_{\rm F}=-1eV+\lambda$ doped cases as a function of photon energy (in units of eV) around the $K$ point. Electron and hole masses are set to be $0.5m_0$ and for two set of parameters, $\beta=0, t_0=2.02eV$ and $\beta=1.77, t_0=1.51$.}
\label{fig1}
\end{figure}

\begin{figure}
\includegraphics[width=1\linewidth]{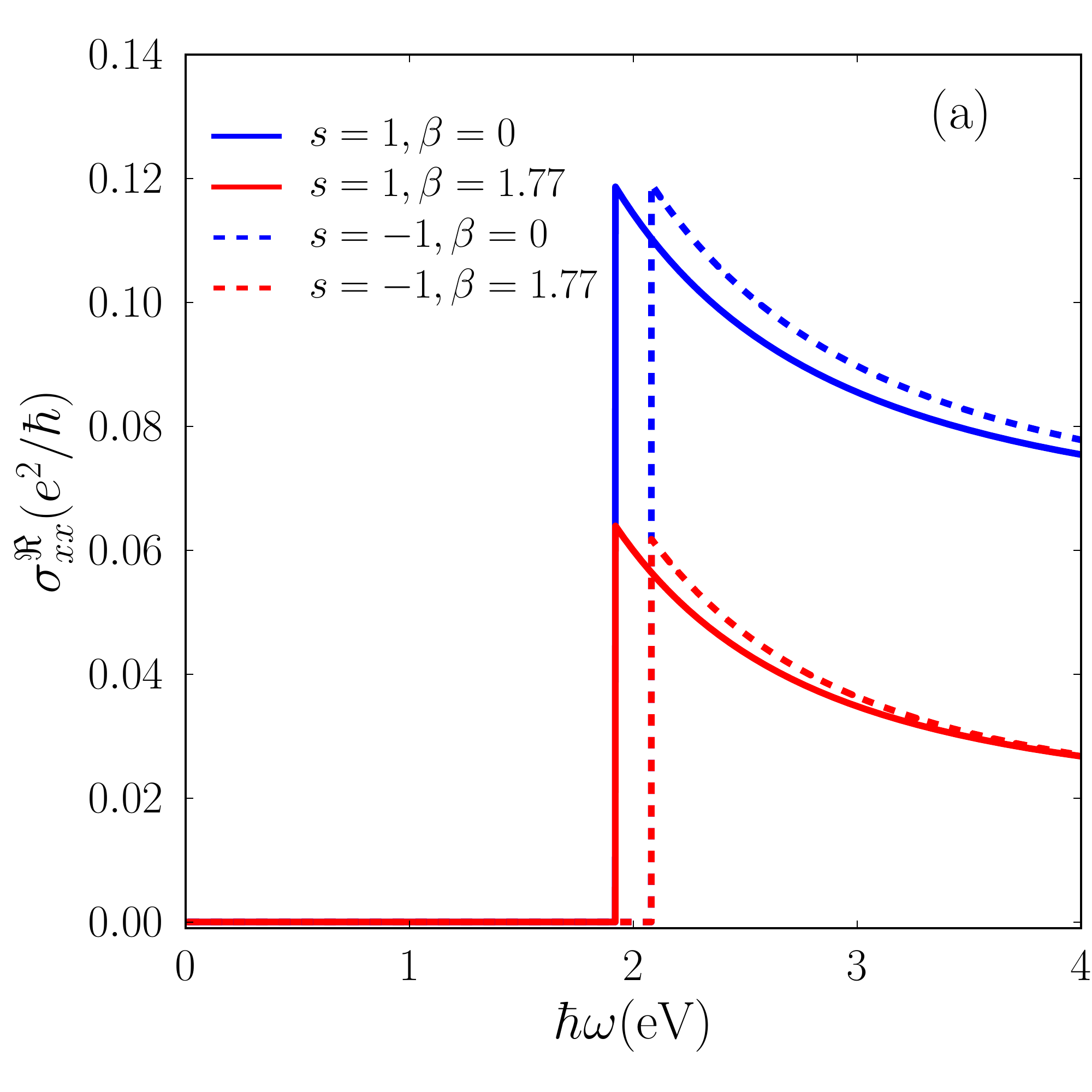}
\includegraphics[width=1\linewidth]{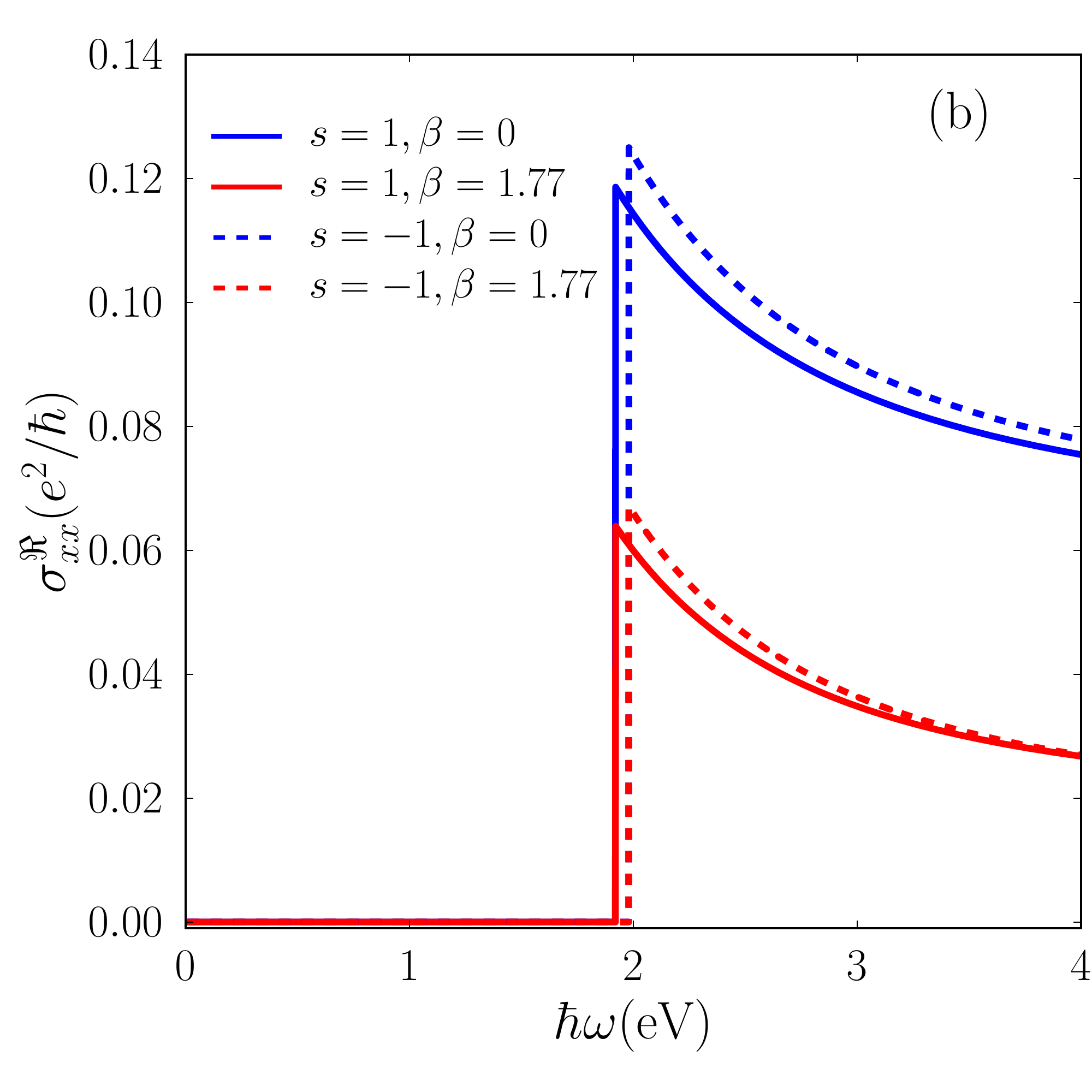}
\caption{(Color online). Real part of the longitudinal conductivity (in units of $e^2/\hbar$) for (a) electron with $\varepsilon_{\rm F}=1eV$ and (b) hole
with $\varepsilon_{\rm F}=-1eV+\lambda$ doped cases as a function of photon energy (in units of eV) around the $K$ point. Electron and hole masses are set to be $0.5m_0$ and for two set of parameters, $\beta=0, t_0=2.02eV$ and $\beta=1.77, t_0=1.51$.}
\label{fig2}
\end{figure}

The real part of the optical Hall and longitudinal conductivities
for the two set of parameters, with and without quadratic terms, are illustrated in
Fig.~\ref{fig1} and Fig.~\ref{fig2} where top and
bottom panels indicate electron and hole doped systems, respectively. The
effect of the mass asymmetry between the effective masses of the
electron and hole ($\alpha$) bands is neglected and it will be discussed
later. It is clear that the quadratic term, $\beta$, causes a reduction of the intensity of the optical Hall conductivity
with no changing of the position of peaks for both electron and
hole doped cases. The position of peaks in the real part of Hall
conductivity is given by
$\hbar\omega=\sqrt{(\Delta-\lambda\tau s)^2+4t^2_0{q_{\rm F
s}}^2}$ for $\beta=0$ case and ${\hbar\omega'}m(q_{\rm F
s}){n(\omega')}^{-1}-2\sqrt{(\Delta'_{\tau  s}+\beta'{q_{\rm F
s}}^2)^2+{q_{\rm F s}}^2}=0~~\text{and}~~
{\hbar\omega'}-2\sqrt{(\Delta'_{\tau  s}+\beta'{q_{\rm F
s}}^2)^2+{q_{\rm F s}}^2}=0$ for each spin component with corresponding Fermi wave
vector $q_{\rm F s}$ and for the case that $\beta\neq0$.
Surprisingly, the last two equations for the later case are simultaneously fulfilled the equation $m(q_{\rm F s})=n(\omega')$ in frequency. In the energy
range shown in the figures, the numerical value of the peak
position for both cases are approximately equal and it indicates
that the position of peaks and steplike configuration
don't change due to the $\beta$ term in a certain Fermi energy. It should be noticed that
the intensity of the real part of $\sigma_{xx}$ decreases with the
quadratic term. Consequently, it indicates that the effective mass approximation
of the Hamiltonian for the ML-MoS$_2$ is not completely valid because two sets of parameters with the same effective masses are
showing distinct results.

\subsection{Mass asymmetry between electron and hole}
\begin{figure}
\includegraphics[width=1\linewidth]{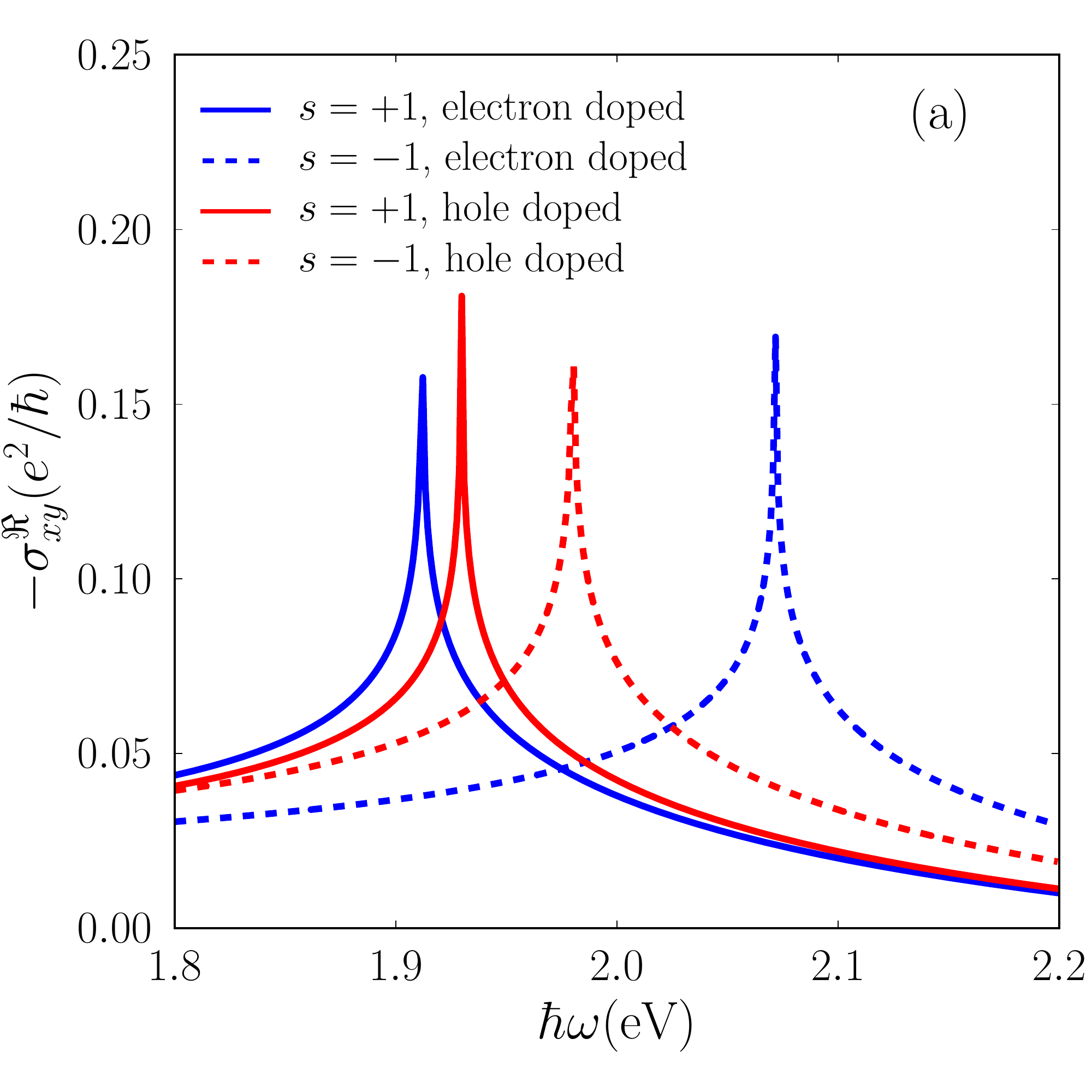}
\includegraphics[width=1\linewidth]{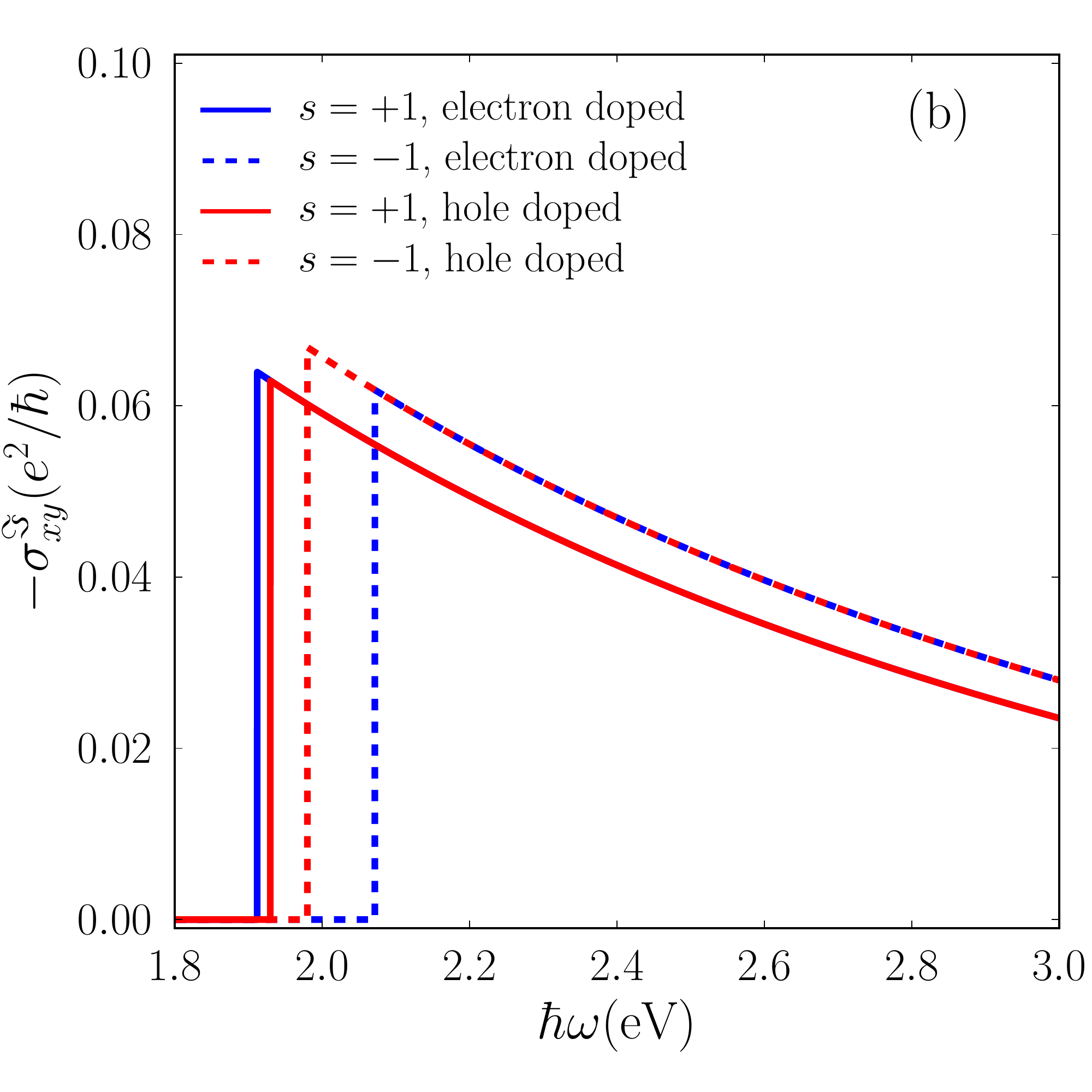}
\caption{(Color online). (a) Real and (b) imaginary parts of the optical Hall conductivity (in units of $e^2/\hbar$) as a function of photon energy (in units of eV) around the $K$ point.
Red (blue) color stands for electron (hole) doped case with $\varepsilon_{\rm F}=1eV$ ($\varepsilon_{\rm F}=-1eV+\lambda$) and solid (dashed) line indicates the spin up (down).}
\label{fig3}
\end{figure}

\begin{figure}
\includegraphics[width=1\linewidth]{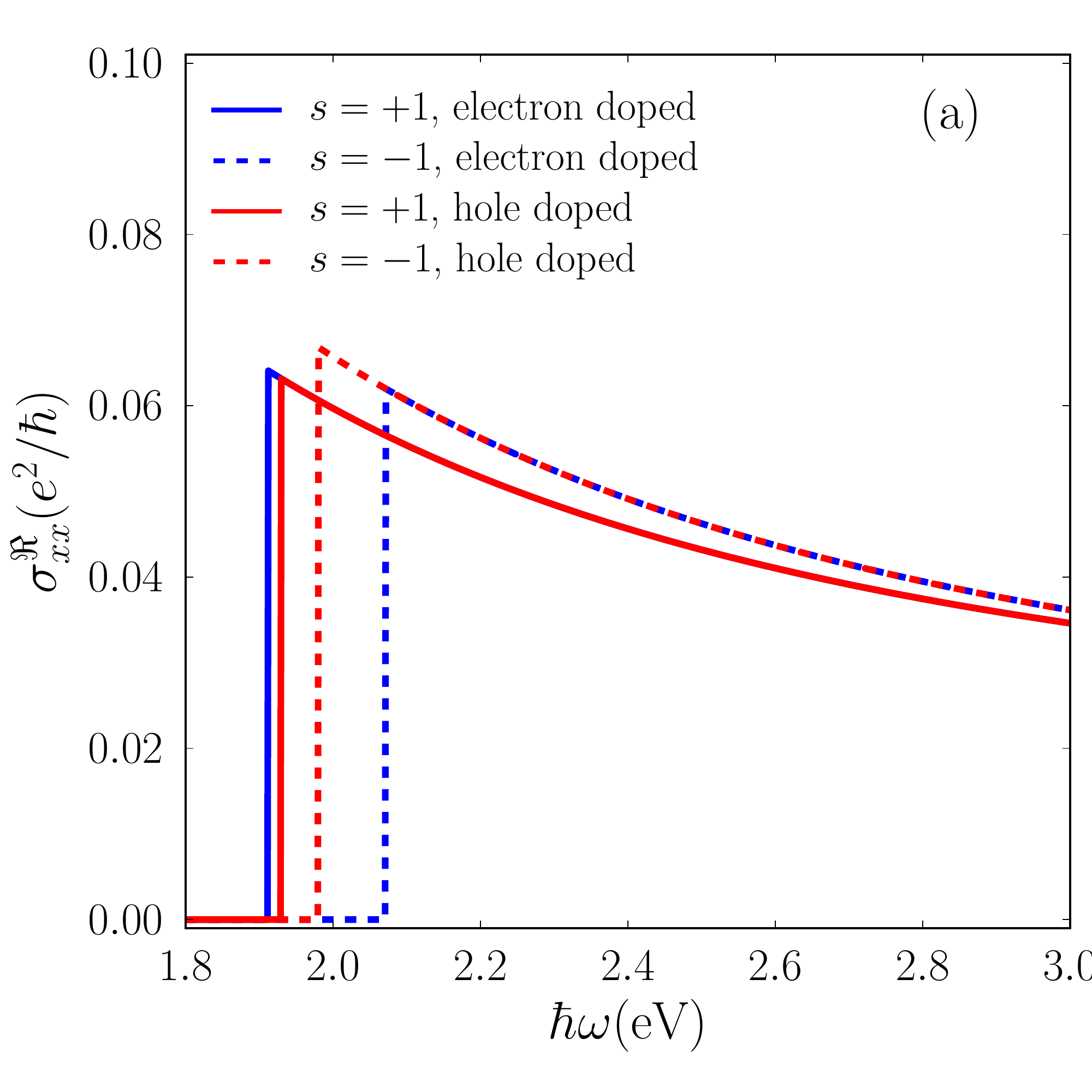}
\includegraphics[width=1\linewidth]{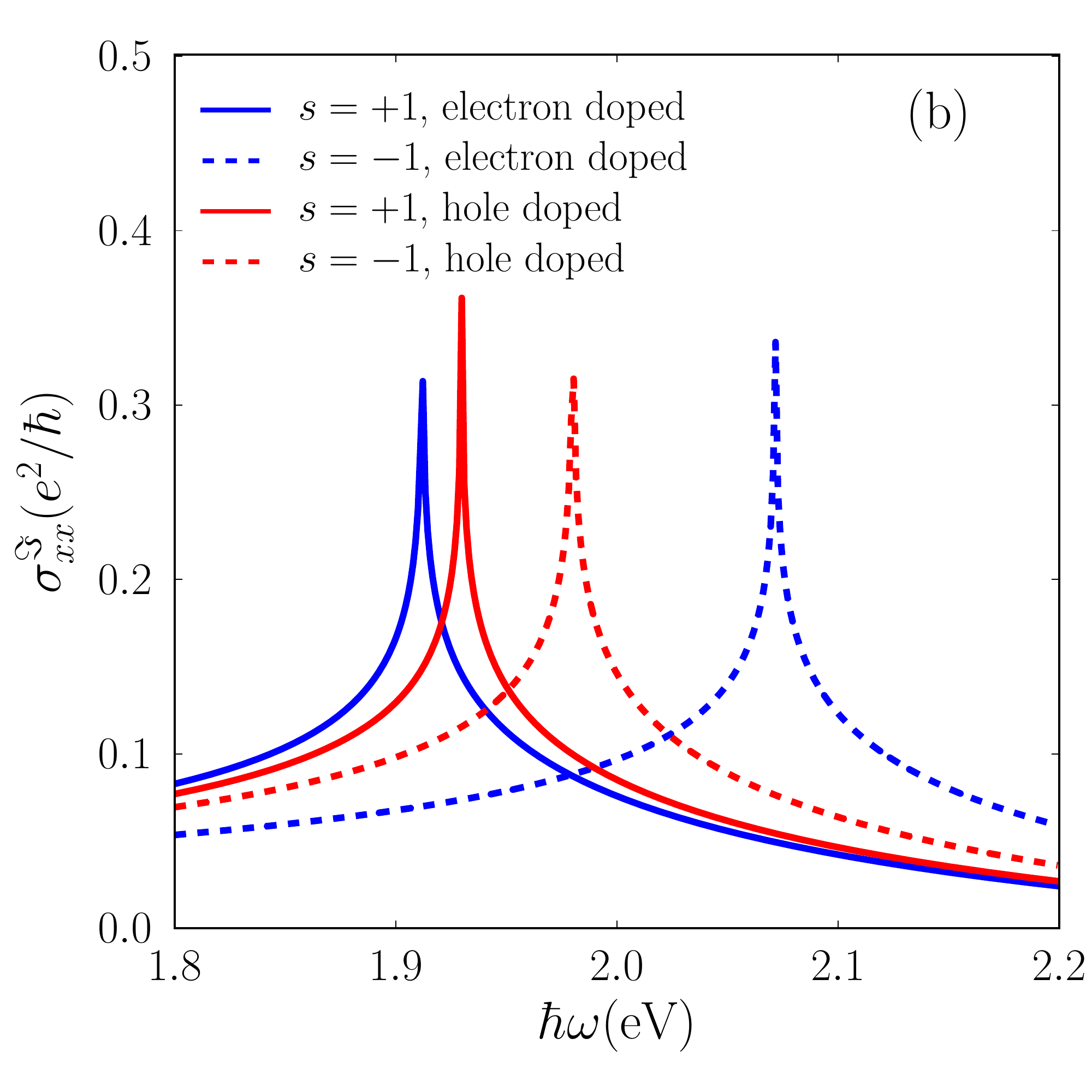}
\caption{(Color online). (a) Real and (b) imaginary parts of the optical longitudinal conductivity (in units of $e^2/\hbar$) as a function of photon energy (in units of eV) around the $K$ point.
Red(blue) color stands for the electron (hole) doped case with $\varepsilon_{\rm F}=1eV$ ($\varepsilon_{\rm F}=-1eV+\lambda$) and the solid (dashed) line indicates the spin up (down).}
\label{fig4}
\end{figure}

In this subsection, we consider the mass asymmetry between electron
and hole bands and then the conductivity of the ML-MoS$_2$ is calculated
for the Hamiltonian given in Eq. (\ref{hamiltonain}). The
results are illustrated in Figs.~\ref{fig3} and \ref{fig4} around the $K$ point. Due to
the mass asymmetry, a small splitting between electron and hole
doped cases takes place in the spin-up component.
On the other hand, there is 
considerable splitting between electron and hole doped cases due to
both spin-orbit coupling and mass asymmetry for the spin-down case. We also note a sharp onset in the imaginary part of the conductivity, minimum energy
associated with the possible interband optical transition.
Moreover, corresponding to the onset in $\sigma^{\Im}_{xy}~(\sigma^{\Re}_{xx})$ where there is a peak in its real (imaginary) part at the same energy as they are related by the Kramers-Kroning relations.

\begin{figure}
\includegraphics[width=1\linewidth]{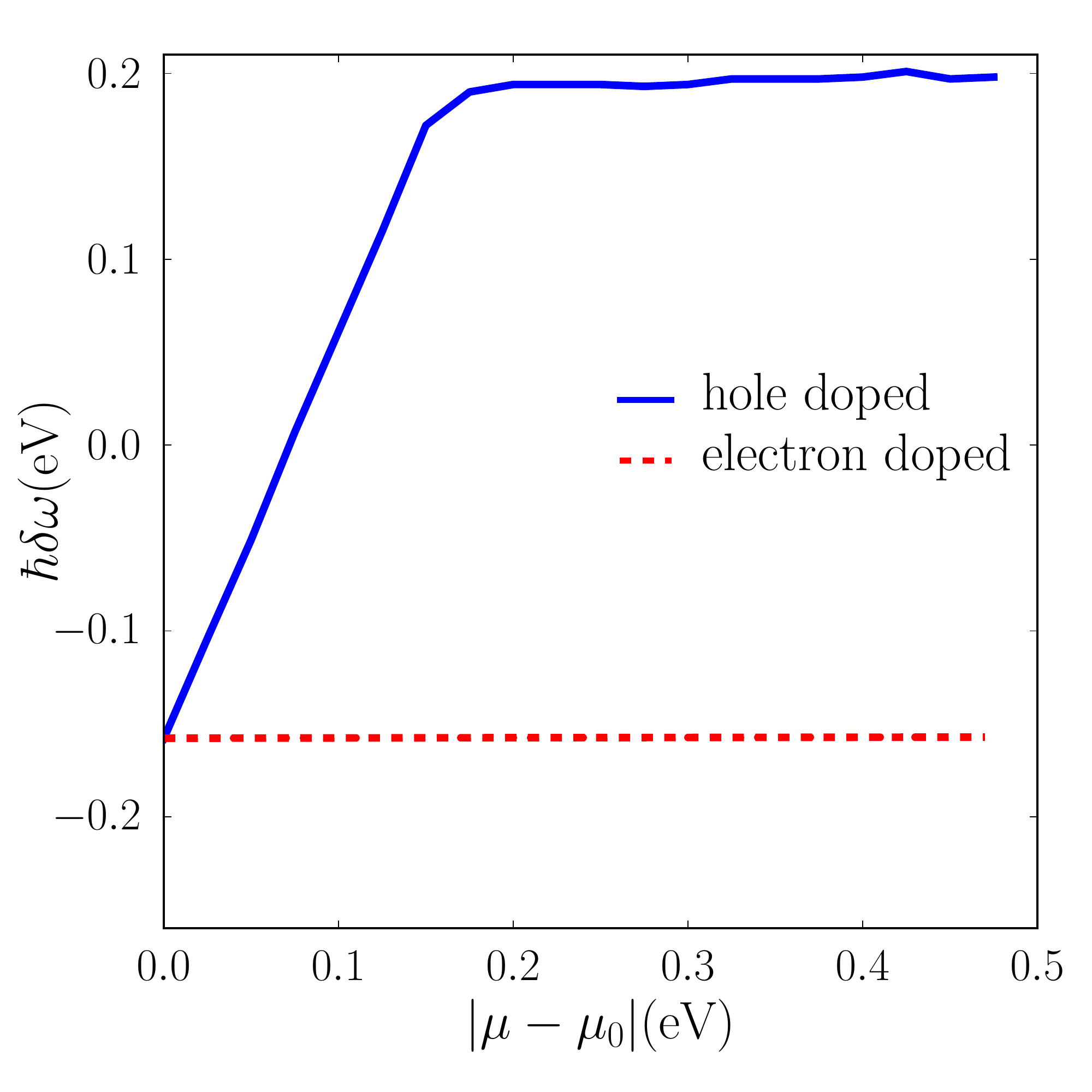}
\caption{(Color online) Difference between the position of the peak in the real part of the Hall conductivity, $\delta\omega=\omega_{\uparrow}-\omega_{\downarrow}$ for the both spin components
for the electron doped case including mass asymmetry as a function of the chemical potential.
Note that $\mu_0$, which is the band edge in the conduction and valence bands, is $0.95$eV and $-0.87$eV for the electron and hole
doped, respectively.}
\label{fig5}
\end{figure}

The position of peaks or steplike configuration of the dynamical conductivity,
can be controlled by the doping rate. Figure.~\ref{fig5} shows the difference between the position of
those peaks, $\delta\omega=\omega_\uparrow-\omega_\downarrow$, around the $K$
point for electron and hole doped cases corresponding to the real part
of the Hall conductivity for each spin component. As it is clearly shown in this figure, $\delta\omega$ increases linearly from a negative value to a positive one
up to a saturation value ($2\lambda)$ for the hole doped case. The linear
part of the result originates from the spin splitting in the
valence band and the fact that there is two fermi wave vectors in which one component spin has zero Fermi wave vector and does not change by increasing the doping rate. Finally, by increasing the Fermi energy, two Fermi wave vectors contribute to the calculations and the position of both peaks move in the same way
and lead to a saturation value
for $\delta\omega$.

\subsection{Circular dichroism and Optical transmittance}

\begin{figure}
\includegraphics[width=1\linewidth]{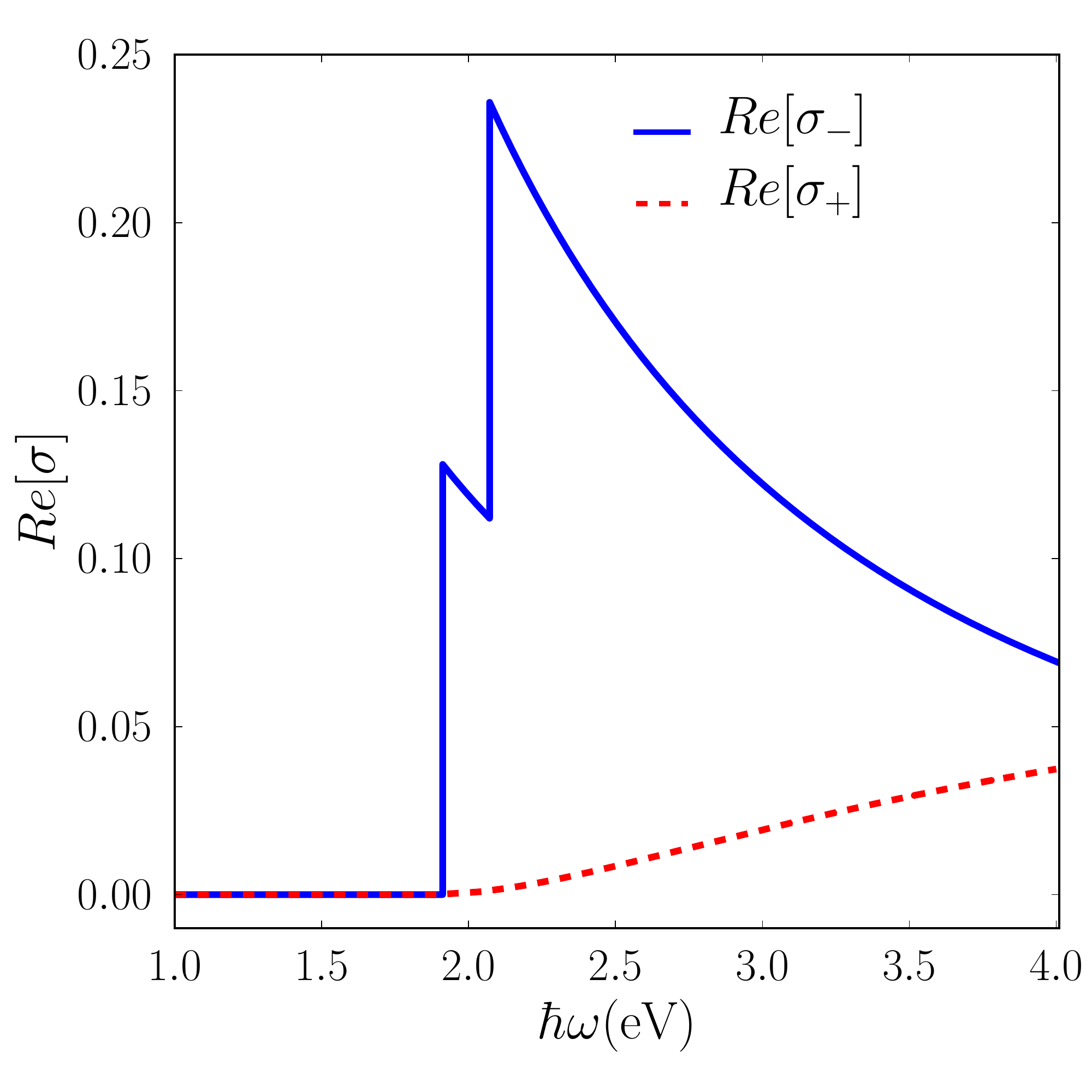}
\caption{(Color online) Real part of the optical conductivity around $K$ point, for left (solid) hand right (dashed) handed light.
It indicates the appearance of the circular dichroism effect for the modified-Dirac equation.
The electron ($\varepsilon_{\rm F}=1eV$) doped case including mass asymmetry.}
\label{fig6}
\end{figure}

One of the main optical properties of the monolayer transition metal
dichalcogenide system is the circular dichroism when it is exposed by a circularly polarized light in which left-
or right-handed light can be absorbed only by $K$ or $K'$ valley and
it makes the material promising for the valleytronic field. This
effect originates from the broken inversion symmetry and it can be
understood by calculating the interband optical selection rule
 ${\cal P_{\pm}}=m_0\langle\psi_c|v_x\pm i v_y|\psi_v\rangle$ for incident right-(+) and left-(-)handed light. The photoluminescence probability for the modified Dirac fermion Hamiltonian is
\begin{eqnarray}
|{\cal P_\pm}|=\frac{m_0 t_0 a_0}{\hbar}(1\pm\tau\frac{d-2b\beta q^2}{\sqrt{d^2+c^2 q^2}})
\end{eqnarray}
where $q^2=q_x^2+q_y^2$. Notice that the mass asymmetry term,
$\alpha$, has no effect on the optical selection rule. The selection rule can simply prove
the circular dichroism in the ML-MoS$_2$. Another approach which helps us to
understand this effect is to calculate the optical conductivity
around the $K$ point of two kinds of light polarizations as
$\sigma_{\pm}=\sum_{s}\{\sigma^{K s}_{xx}\pm \sigma^{K s}_{xy}\}$ which has
been calculated by using the Dirac-like model~\cite{yao08,Li12} and now, we modify
that by using the modified-Dirac Hamiltonian. Figure.~\ref{fig6} shows the coupling of the light and valleys.
Note that $\Re e [\sigma_{-}]$ is large and comparable in size for either spin up or down while $\Re e [\sigma_{+}]$
is small in comparison. The valley around the $K$
point can couple only to the left-handed light and this effect is
washed up by increasing the frequency of the light and the result is
in good agreement with recent experimental measurements~\cite{mak12}.

\begin{figure}
\includegraphics[width=1\linewidth]{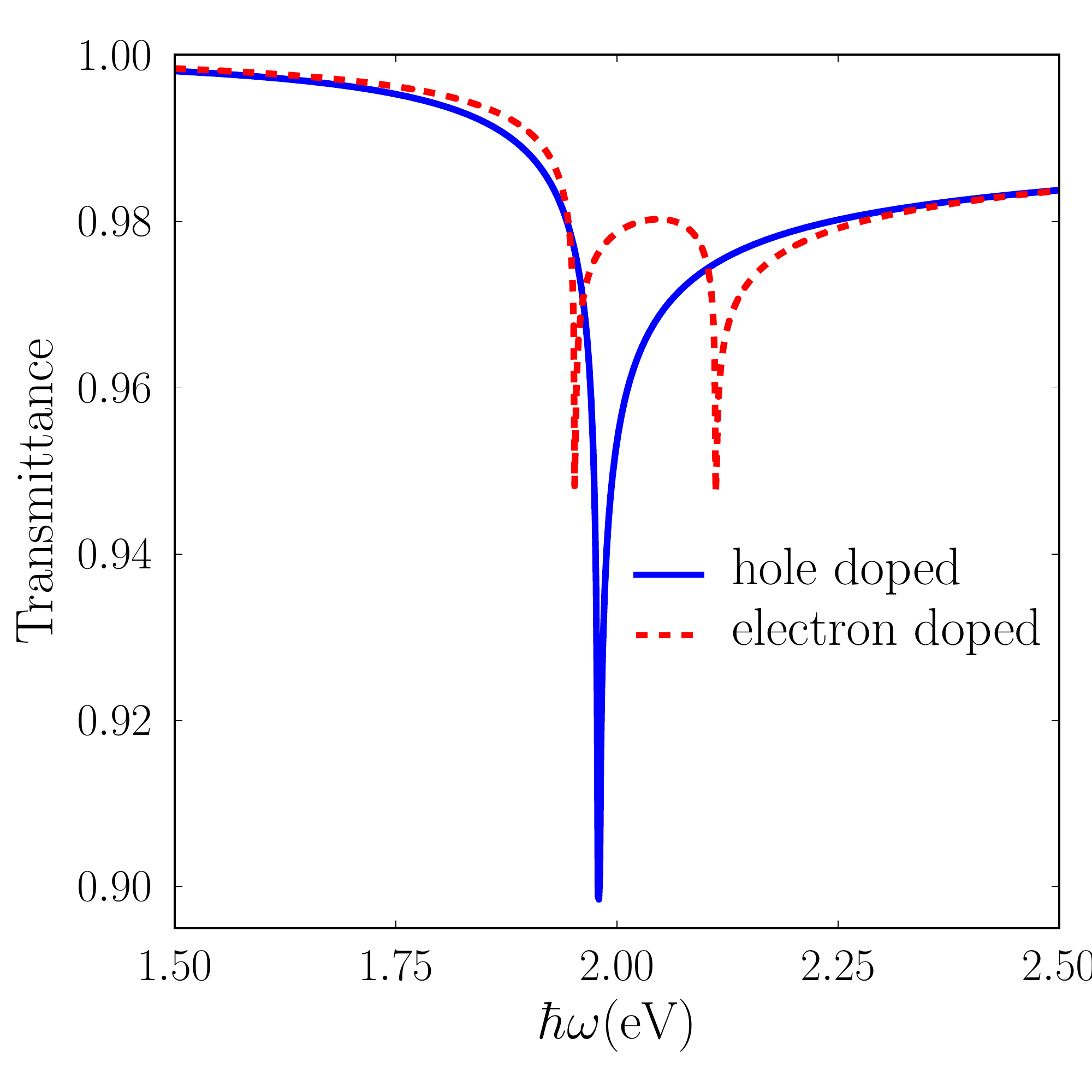}
\caption{(Color online) Optical transmittance in a finite frequency for the electron ($\varepsilon_{\rm F}=1eV$) and hole
($\varepsilon_{\rm F}=-1eV+\lambda$) doped cases including mass asymmetry.}
\label{fig7}
\end{figure}

Furthermore, the optical transmittance is an important physical quantity and it can be evaluated stemming from the conductivity. The optical transmittance of a free standing thin film
exposed by a linear polarized light is given by~\cite{Ferreira11}
\begin{eqnarray}
T(\omega)=\frac{1}{2}\{|\frac{2}{2+Z_0\sigma_{+}(\omega)}|^2+|\frac{2}{2+Z_0\sigma_{-}(\omega)}|^2\}
\end{eqnarray}
where $Z_0=376.73\Omega$ and
$\sigma_{\pm}(\omega)=\sigma_{xx}(\omega)\pm i\sigma_{xy}$ are the
vacuum impedance and the optical conductivity of the thin film,
respectively. For the ML-MoS$_2$ case, the total Hall conductivity
in the presence of the time reversal symmetry is zero and the
total longitudinal conductivity is given by
$\sigma_{xx}=2(\sigma_{xx}^{K\uparrow}+\sigma_{xx}^{K\downarrow})$.
The optical transmittance of the multilayer of MoS$_2$ systems has been
recently measured~\cite{Gomez13} and it is about $94.5\%$ for each
layer in the optical frequency range. The optical transmittance of
the ML-MoS$_2$ is displayed in Fig.~\ref{fig7} for both
electron and hole doped cases using the numerical value defined as
$set_0$. The result shows that the optical transmittance is about $98\%$ for the frequency range in which
both spin components are active for giving response to the
incident light. Importantly, for the electron dope case,
there are two minimums with distance about $0.16\text{eV}/\hbar$ in frequency which mostly indicates the spin-orbit
splitting ($2\lambda$) in the valence band and it is consistent with the results illustrated in Fig.~\ref{fig5}. The optical transmittance for electron doped case is about $98\%$ in all frequency range.
Moreover, for the hole dope case as it is shown in Fig.~\ref{fig5}, the optical transmittance changes by tuning doping rate.
Interestingly, at $\mu=-0.942$eV the difference between the position of peaks of two spin components, $\delta \omega$ is approximately zero.
Consequently, the total optical conductivity enhances in this resonating doping rate which has significant effect on the optical transmittance of
the system where the transmittance decreases and particularly reaches to a value less than $90\%$ at the resonance frequency when $\delta \omega\simeq 0$. Our numerical calculations show that the hole doped ML-MoS$_2$ is darker than the electron doped one specially close to the resonance frequency.
Furthermore, this feature provides an opportunity with measuring the spin-orbit coupling by an optical transmittance measurement.

\subsection{Optical response in the non-trivial phase}\label{sec:TI}

The modified-Dirac Hamiltonian shows a non-trivial phase when
$\beta\Delta<0$ and it has been numerically shown that in this
phase a light matter interaction enhances due to the change of the
parabolic band dispersion into the shape of a
Mexican-hat with two extrema~\cite{peres13}. To fulfill such a
band dispersion, a negative value $\beta\Delta$ with a large absolute value is required
and it is accessible for an ultrathin film of the topological
insulator. The sign and the absolute values of the parameters
can be manipulated by the thickness of the thin film, while
in the case of the ML-MoS$_2$, to the best of our knowledge, it is barely
possible to create a Mexican hat like dispersion relation even for the
model Hamiltonian with a non-trivial
topology phase~\cite{Kormanyos13,Liu13}. In this case, we plot the
optical Hall and longitudinal conductivities of the UTF-TI in its
trivial and non-trivial phases. In the UTF-TI~\cite{zhang09} system, in which only
in-plan components of momentum are relevant, one can find the
Hamiltonian given by Eq.~(\ref{hti}) where the numerical value of the
model parameters depends on the thickness of the thin
films~\cite{Lu10,Shan10}. We consider three
different thicknesses for which three sets of
parameters~\cite{Lu10} are listed in Table I. We also neglect the value of $\epsilon_0$ which is just a
constant shift in the energy.
\begin{table}[ht]
\caption{Numerical parameter for the ultra thin film of a
topological insulator.\cite{Lu10}} \centering
\begin{tabular}{|c|c|c|c|c|c|}
\hline
L(\AA)&$\Delta$(eV)&$t_0$(eV)&$\alpha$& $\beta$\\
\hline
20          & 0.14 & -2.22 & -1.05 & 23.67\\
\hline
25          & 0.0 & -2.21 & -2.37 & 18.41   \\
\hline
32          &-0.04 & -2.20 & -3.94 & 6.31 \\
\hline
\end{tabular}
\label{table:ti}
\end{table}
As it can be seen from table I, a sample with $L=20\AA$  or
$L=32\AA$ indicates the trivial or non-trivial phases,
respectively. However for a sample with $L=25\AA$ the energy gap vanishes and thus at critical thickness, $L=25\AA$,
the trivial to non-trivial phase transition takes place. Hereafter, we call that a phase boundary.

\begin{figure}
\includegraphics[width=1\linewidth]{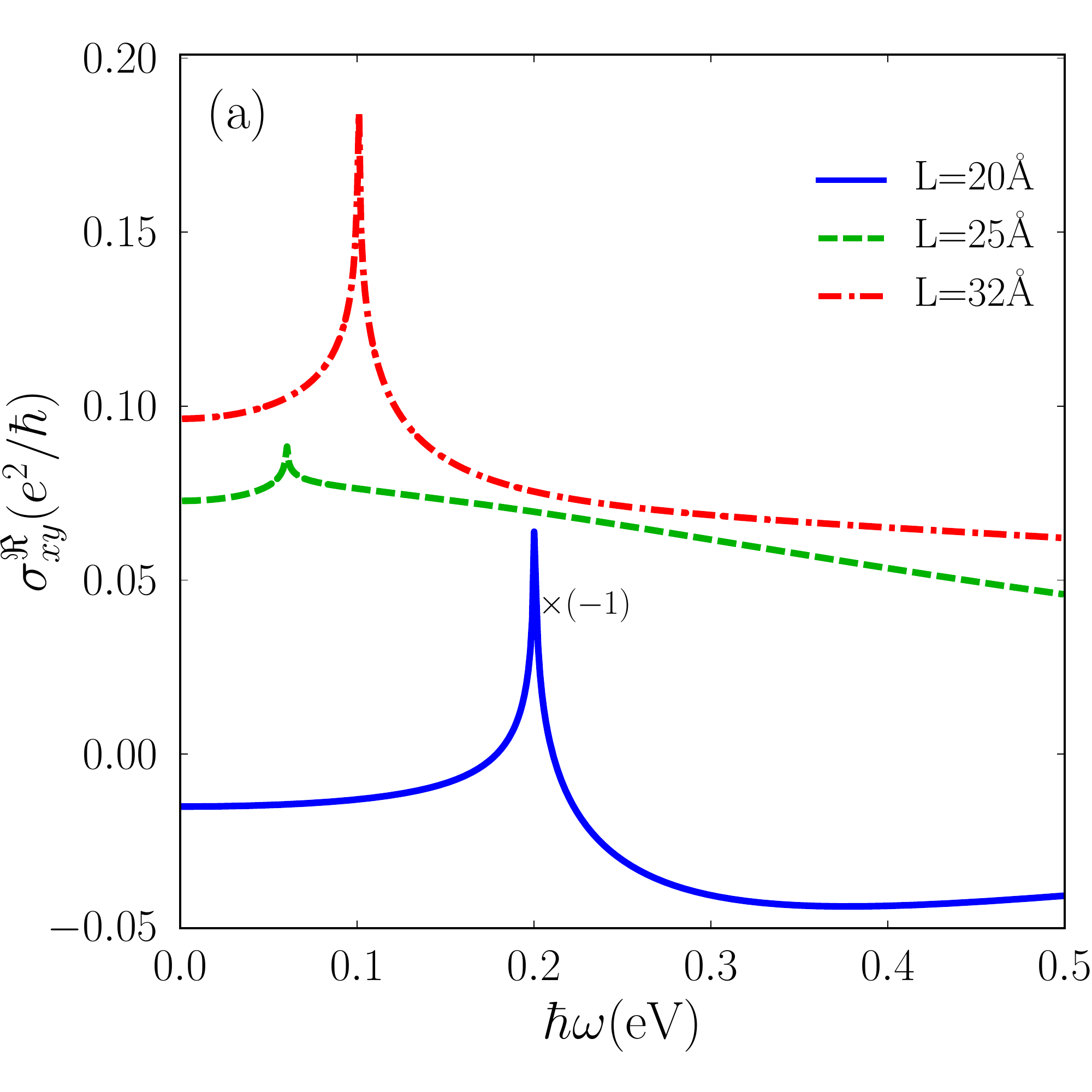}
\includegraphics[width=1\linewidth]{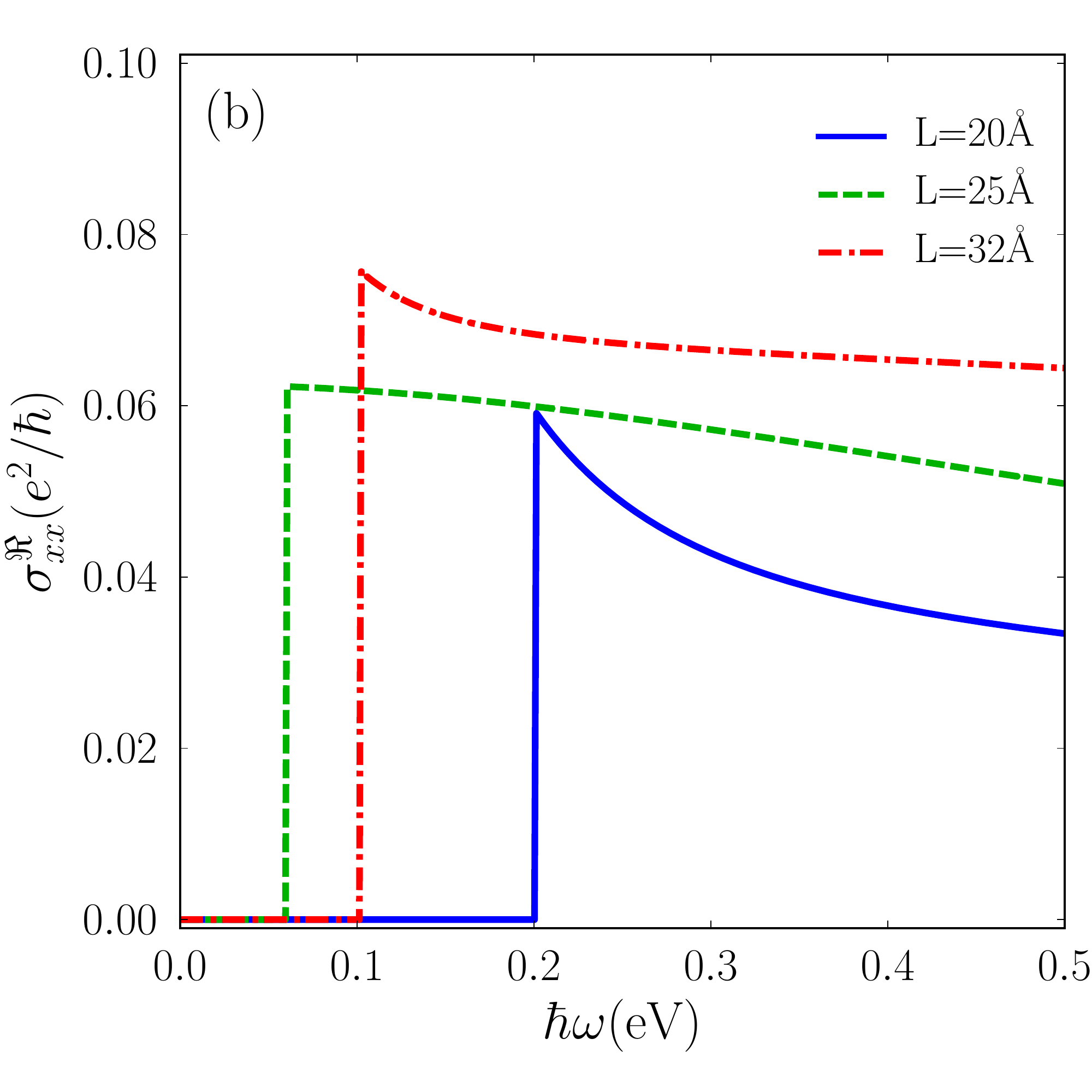}
\caption{(Color online) Real part of the Hall (a) and longitudinal (b) conductivity for $\tau=1$ and different values of film thickness.
It is clear that in the non-trivial phase the optical response of the system is stronger than that of its trivial one. The Fermi energy is $\varepsilon_{\rm F}=|\Delta|/2+0.03$eV.}
\label{fig8}
\end{figure}

\begin{figure}
\includegraphics[width=1\linewidth]{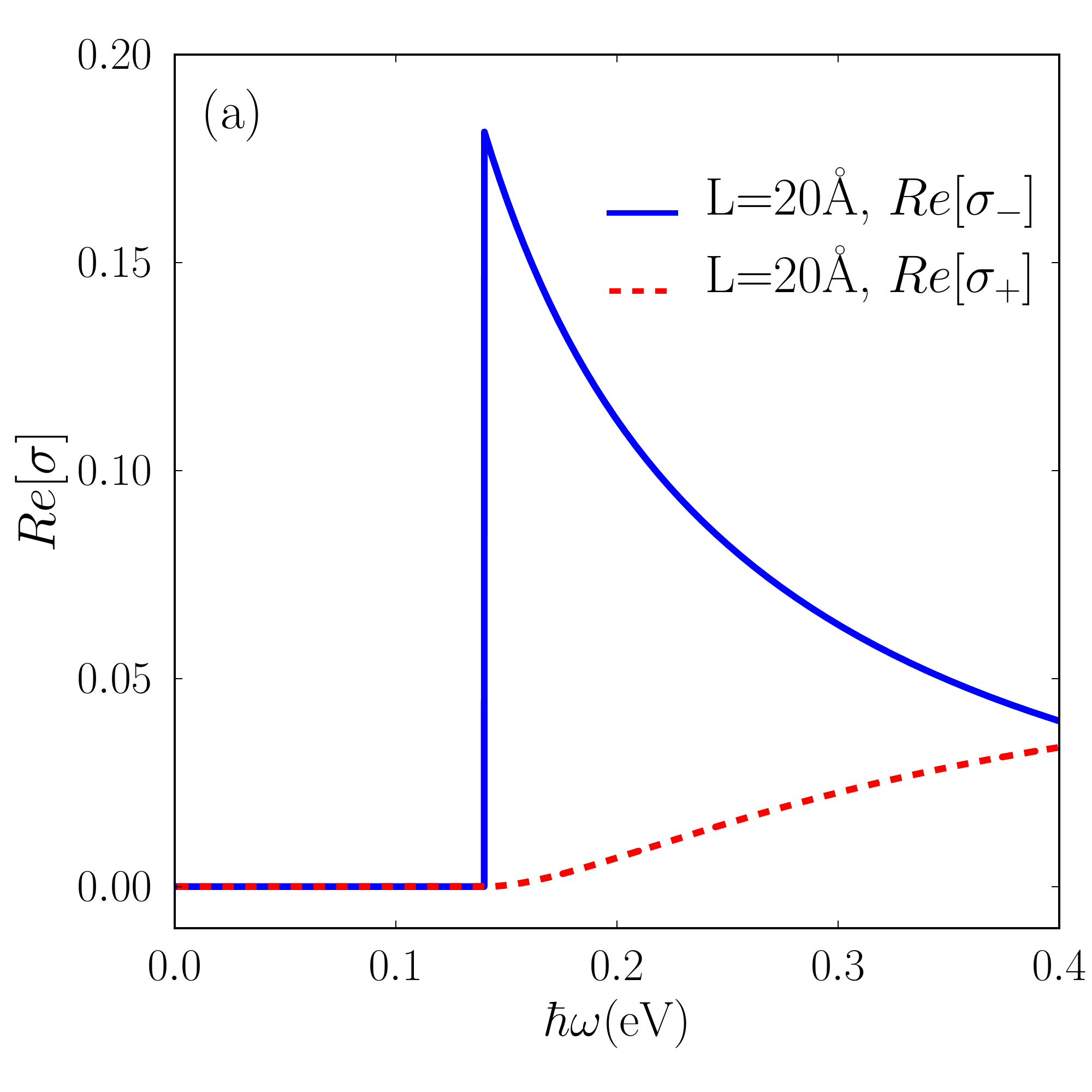}
\includegraphics[width=1\linewidth]{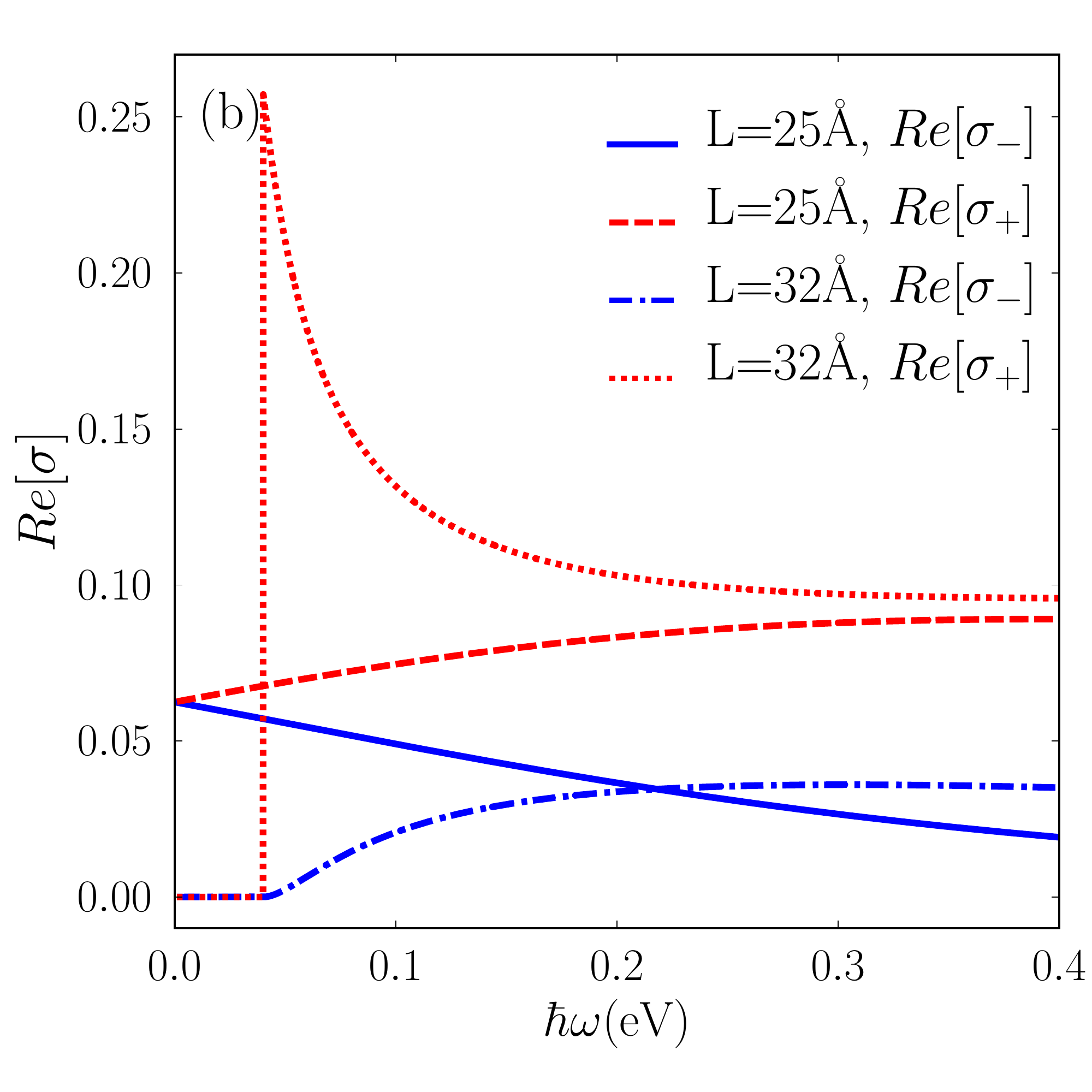}
\caption{(Color online) Circular dichroism effect for different values of the thickness. The real part of the optical conductivity
around the $K$ point is shown for (a) $L=20\AA$ and (b) $L=25\AA$ and $31\AA$. The Fermi energy is $\varepsilon_{\rm F}=|\Delta|/2+0.03$eV.}
\label{fig9}
\end{figure}

Now, we calculate the real part of the Hall and longitudinal
conductivities for $\tau=+$ and the results are illustrated in
Fig.~\ref{fig8}. It shows that the
conductivity enhances in the non-trivial phase which is consistent with previous numerical
work.~\cite{peres13} More interestingly, we are now showing that the Hall
conductivity changes sign through changing the thickness and it
is very important in the circular dichroism effect. This
changing of the sign means a different helicity of the light can be
coupled to the system. It is worth mentioning that the
circular dichroism effect on the electronic system governing
modified-Dirac Hamiltonian is also possible when energy gap
is zero~\cite{yao08,Li12}. The selection rule
equation reads as $|{\cal
P_\pm}|=\frac{m_0 t_0 a_0}{\hbar}(1\mp\tau{b\beta
q}/{\sqrt{(b\beta q)^2+c^2}})$ for the case of zero gap. This expression indicates that the circular polarization is achievable away from the
$\Gamma$ point even in the
absence of the energy gap. It might be emphasized that the peak in the optical conductivity at zero energy gap originates from a non-zero Fermi energy in which the low energy part of phase space is no longer available for a photon absorbtion process~\cite{jahn} based on the Pauli exclusion principle. More precisely, there is a peak at energy point $\hbar\omega\approx2\varepsilon_{\rm F}$ in the topological insulator case and it can be seen from Eqs. (\ref{sxy}) and (\ref{sxx}). Therefore, the peak disappears at zero Fermi energy for a gapless system. In Fig.~\ref{fig9}, we show the
optical conductivity for the two helicities of light for $\tau=+$.
The results show that the circular polarization
changes sign for negative value of the gap and it gets more
strength in the non-trivial phase rather than the trivial phase.

\section{summary}\label{sec:summary}

We have analytically calculated the intrinsic
conductivity of the electronic systems which govern a
modified-Dirac Hamiltonian by using the Kubo formula. We have studied the effect
of the quadratic term in momentum, $\beta$, which has been recently
predicted, and found the different
optical responses. This discrepancy originates from the different
topological structures of the systems. Our calculations show that the
$\beta$-term has no effect on the position of the peak of the
optical conductivity but it has considerable effect on its magnitude.
Therefore, it shows that the same effective mass approximation
for electron and hole bands for monolayer MoS$_2$ can not fully describe the optical properties. The
effect of the strong spin-orbit interaction can be traced by the
difference of the energy interval between the position of the peak
in the optical conductivity for the two spin components in
electron and hole doped cases. We have shown that this interval
for the electron doped case is approximately constant while for the hole
doped case, it increases from a negative value to a positive one, and then it
increases linearly up to a saturation value. The effect of the mass
asymmetry in monolayer MoS$_2$ induces a small splitting between
the conductivity spectrum for the electron and hole doped cases.
The circular dichroism effect is investigated for the
modified-Dirac Hamiltonian of the monolayer MoS$_2$ by
calculating the selection rule and the optical conductivity. We have also obtained the
optical transmittance of the monolayer MoS$_2$ for the hole and
electron doped cases and the results show that the valence band spin splitting has considerable
effect on the intensity of the transmittance.

We have also studied the effect of the quantum phase transition,
which occurs owing to the reducing of the thickness, on the optical
conductivity of the thin film of the topological insulator. We
have shown that at the phase boundary, when the energy gap is zero, the
diagonal quadratic term plays a significant role on the optical conductivity
and selection rule. Moreover, we have illustrated that the optical
response enhances and the optical Hall conductivity changes sign in
the non-trivial phase (QSH) and the phase boundary.

\begin{acknowledgments}

R. A. would like to thank the Institute for Material Research in Tohoku University for its hospitality during the period when the last
part of this work was carried out.

\end{acknowledgments}

\appendix
\numberwithin{equation}{section}
\section{}
In this appendix, the details of the calculations deriving
Eq.~(\ref{conductivity}) are presented. Since
$\langle\psi_c|\psi_v\rangle=0$, we get
\begin{eqnarray}
\langle\psi_c|\hbar v_x|\psi_v\rangle&=&c\langle\psi_c|\sigma_x|\psi_v\rangle +2b\beta q_x\langle\psi_c|\sigma_z|\psi_v\rangle\nonumber\\
\langle\psi_v|\hbar v_y|\psi_c\rangle&=&c\langle\psi_v|\sigma_y|\psi_c\rangle +2b\beta q_y\langle\psi_v|\sigma_z|\psi_c\rangle
\end{eqnarray}
owing to the fact that the mass asymmetry parameter $\alpha$ plays no role in
the velocity matrix elements. Using $h_ch_v=-c^2q^2$ we have
\begin{eqnarray}
\langle\psi_c|\sigma_x|\psi_v\rangle&=&\frac{-c}{D_cD_v}[qh_v+q^\ast h_c]\nonumber\\
\langle\psi_v|\sigma_y|\psi_c\rangle&=&\frac{ic}{D_cD_v}[qh_c-q^\ast h_v]\nonumber\\
\langle\psi_c|\sigma_z|\psi_v\rangle&=&\langle\psi_v|\sigma_z|\psi_c\rangle=\frac{2c^2q^2}{D_cD_v}
\end{eqnarray}
In this case
\begin{eqnarray}
\langle\psi_c|\hbar v_x|\psi_v\rangle&=&\frac{c^2}{D_cD_v}\{-[qh_v+q^\ast h_c]+4b\beta q_x q^2\}\nonumber\\
\langle\psi_v|\hbar v_y|\psi_c\rangle&=&\frac{c^2}{D_cD_v}\{i[qh_c-q^\ast h_v]+4b\beta q_y q^2\}
\end{eqnarray}
Consequently, the product of the velocity matrix elements are
\begin{eqnarray}\label{eq1}
&&\langle\psi_c|\hbar v_x |\psi_v\rangle\langle\psi_v|\hbar v_y |\psi_c\rangle=\frac{c^4}{(D_cD_v)^2}\{-i(qh_v+q^\ast h_c)(q h_c-q^\ast h_v)\nonumber\\&&+
(4b\beta q^2)^2q_xq_y+4b\beta q^2(-q_y(qh_v+q^\ast h_c)+iq_x(q h_c-q^\ast h_v))\}\nonumber\\&&
\langle\psi_c|\hbar v_x |\psi_v\rangle\langle\psi_v|\hbar v_x |\psi_c\rangle=\frac{c^4}{(D_cD_v)^2}\{|qh_v+q^\ast h_c|^2\nonumber\\&&+(4bq_x\beta q^2)^2-4bq_x\beta q^2(qh_v+q^\ast h_c+q^\ast h_v+q h_c)\}
\end{eqnarray}
Using $\tan\phi=q_y/q_x$, one can find
\begin{eqnarray}\label{eq2}
(q h_v+q^\ast h_c)(q h_c-q^\ast h_v)&=&-2ic^2q^4\sin2\phi\nonumber\\&-&4q^2d\sqrt{d^2+c^2q^2}\nonumber\\
-q_y(q h_v+q^\ast h_c)+iq_x(q h_c-q^\ast h_v)&=&2q^2[-i\sqrt{d^2+c^2q^2}\nonumber\\&+&d\sin2\phi]\nonumber\\
qh_v+q^\ast h_c+q^\ast h_v+q h_c&=&4qd\cos\phi\nonumber\\
|qh_v+q^\ast h_c|^2&=&4q^2(d^2+c^2q^2{\sin\phi}^2)\nonumber\\
(D_cD_v)^2&=&4c^2q^2[d^2+c^2q^2]
\end{eqnarray}
After substituting Eq.~(\ref{eq2}) into Eq.~(\ref{eq1}), we get
\begin{eqnarray}
\langle\psi_c|\hbar v_x |\psi_v\rangle\langle\psi_v|\hbar v_y |\psi_c\rangle&=&\frac{c^2q^2\sin2\phi}{d^2+c^2q^2}\{-\frac{c^2}{2}+2b\beta(b\beta q^2+d)\}\nonumber\\&+&i\frac{c^2}{\sqrt{d^2+c^2q^2}}\{d-2b\beta q^2\}\nonumber\\
\langle\psi_c|\hbar v_x |\psi_v\rangle\langle\psi_v|\hbar v_x |\psi_c\rangle&=&c^2-\frac{c^2q^2{\cos\phi}^2}{d^2+c^2q^2}\{c^2+2b\beta(a_1-a_2)\}\nonumber\\
\end{eqnarray}
Using $\int{d\phi \sin2\phi}=0,\int{d\phi{\cos\phi}^2}=\pi$, one can find
\begin{widetext}
\begin{eqnarray}
&&\sigma_{xy}=\frac{e^2}{h}\int{qdq\frac{f(\varepsilon_c)-f(\varepsilon_v)}{\varepsilon_c-\varepsilon_v}\times\{\frac{c^2}{\sqrt{d^2+c^2q^2}}(d-2b\beta q^2)\}
\{\frac{1}{\hbar\omega+\varepsilon_c-\varepsilon_v+i0^+}
-\frac{1}{\hbar\omega+\varepsilon_v-\varepsilon_c+i0^+}\}}\nonumber\\&&
\sigma_{xx}=-i\frac{e^2}{h}\int{qdq\frac{f(\varepsilon_c)-f(\varepsilon_v)}{\varepsilon_c-\varepsilon_v}\times\{ c^2-\frac{c^2q^2}{d^2+c^2q^2}[\frac{c^2}{2}+b\beta(a_1-a_2)]\}
\{\frac{1}{\hbar\omega+\varepsilon_c-\varepsilon_v+i0^+}
+\frac{1}{\hbar\omega+\varepsilon_v-\varepsilon_c+i0^+}\}}\nonumber\\
\end{eqnarray}
\end{widetext}
Using $(x+i0^+)^{-1}=\mathbb{P} x^{-1}-i\pi\delta(x)$ where
$\mathbb{P}$ stands for principal value, it is easy to show that
the real and imaginary parts of diagonal and off-diagonal
components of the conductivity tensor read as below
\begin{widetext}
\begin{eqnarray}
&&\sigma^{\Re}_{xy}=\frac{2e^2}{h}\int{qdq (f(\varepsilon_c)-f(\varepsilon_v))\times\{\frac{c^2}{\sqrt{d^2+c^2q^2}}(d-2b\beta q^2)\}
\{\mathbb{P}\frac{-1}{(\hbar\omega)^2-(\varepsilon_c-\varepsilon_v)^2}\}}\nonumber\\&&
\sigma^{\Im}_{xy}=\frac{\pi e^2}{h}\int{qdq \frac{f(\varepsilon_c)-f(\varepsilon_v)}{\varepsilon_c-\varepsilon_v}
\times\{\frac{c^2}{\sqrt{d^2+c^2q^2}}(d-2b\beta q^2)\}
\{\delta(\hbar\omega+\varepsilon_v-\varepsilon_c)-\delta(\hbar\omega+\varepsilon_c-\varepsilon_v)\}}\nonumber\\&&
\sigma^{\Im}_{xx}=-\frac{2e^2}{h}\hbar\omega\int{qdq \frac{f(\varepsilon_c)-f(\varepsilon_v)}{\varepsilon_c-\varepsilon_v}\times\{ c^2-\frac{c^2q^2}{d^2+c^2q^2}[\frac{c^2}{2}+b\beta(a_1-a_2)]\}
\{\mathbb{P}\frac{-1}{(\hbar\omega)^2-(\varepsilon_c-\varepsilon_v)^2}\}}\nonumber\\&&
\sigma^{\Re}_{xx}=-\frac{\pi e^2}{h}\int{qdq \frac{f(\varepsilon_c)-f(\varepsilon_v)}{\varepsilon_c-\varepsilon_v}
\times\{ c^2-\frac{c^2q^2}{d^2+c^2q^2}[\frac{c^2}{2}+b\beta(a_1-a_2)]\}
\{\delta(\hbar\omega+\varepsilon_v-\varepsilon_c)+\delta(\hbar\omega+\varepsilon_c-\varepsilon_v)\}}
\end{eqnarray}
\end{widetext}
To find the conductivity around $K'$ point we must implement the
following changes: $p_x\rightarrow -p_x$ and $\lambda\rightarrow
-\lambda$. Using these transformations, the velocity matrix
elements around the $K'$ point can be calculated by taking
complex conjugation of the corresponding results around the $K$
point. Moreover, according to the following dimensionless
parameters, $\varepsilon_c-\varepsilon_v=2\sqrt{d^2+c^2q^2}$,
and thus $\delta(\hbar\omega+\varepsilon_c-\varepsilon_v)\rightarrow0$ for
positive frequency in absorbtion process. Thus Eq.~(\ref{conductivity}) for the dynamical transverse and
longitudinal conductivity is obtained.

\section{}
In this appendix, the details of calculations for some integrals which appear in
our model are presented. Using new variables $y=\beta'q^2+\Delta'_{\tau s}+(2\beta')^{-1}$
and $a^2=\Delta'_{\tau s}/\beta'+(4\beta'^2)^{-1}$, it is easy to
show that $(\Delta'_{\tau s}+\beta'q^2)^2+q^2=y^2-a^2$ and we have
\begin{eqnarray}
G_{\tau s}(\omega,q)&=&\frac{1}{\beta'}\{(2\Delta'_{\tau s}+\frac{1}{2\beta'})I_1-I_2\}\nonumber\\
H_{\tau s}(\omega,q)&=&\frac{\hbar\omega'}{\beta'}\{I_1-(2\Delta'_{\tau s}+\frac{1}{2\beta'})I_3\nonumber\\&+&(2\Delta'_{\tau s}+\frac{1}{2\beta'})(\Delta'_{\tau s}+\frac{1}{2\beta'})I_4\}
\end{eqnarray}
where $I_1, I_2, I_3$, and $I_4$ are given by
\begin{eqnarray}
I_1&=&\int{\mathbb{P}\frac{dy}{\sqrt{y^2-a^2}[4(y^2-a^2)-(\hbar\omega')^2]}}\nonumber\\
I_2&=&\int{\mathbb{P}\frac{y dy}{\sqrt{y^2-a^2}[4(y^2-a^2)-(\hbar\omega')^2]}}\nonumber\\
I_3&=&\int{\mathbb{P}\frac{y dy}{(y^2-a^2)^{\frac{3}{2}}[4(y^2-a^2)-(\hbar\omega')^2]}}\nonumber\\
I_4&=&\int{\mathbb{P}\frac{dy}{(y^2-a^2)^{\frac{3}{2}}[4(y^2-a^2)-(\hbar\omega')^2]}}
\end{eqnarray}
$I_1$ and $I_4$ can be calculated by defining $u$ as a new
variable where $y=\frac{a}{\sqrt{1-u^2}}$ and it
leads to \begin{eqnarray}
I_1&=&\frac{1}{2\hbar\omega'\sqrt{4a^2+(\hbar\omega')^2}}\ln|\frac{u-\frac{\hbar\omega'}{\sqrt{4a^2+(\hbar\omega')^2}}}{u+\frac{\hbar\omega'}{\sqrt{4a^2+(\hbar\omega')^2}}}|\nonumber\\
I_4&=&\frac{1}{a^2}\{-I_1+\frac{1}{(\hbar\omega')^2u}+\frac{\sqrt{4a^2+(\hbar\omega')^2}}{(\hbar\omega')^3}\nonumber\\&\times&\ln|\frac{u-\frac{\hbar\omega'}{\sqrt{4a^2+(\hbar\omega')^2}}}{u+\frac{\hbar\omega'}{\sqrt{4a^2+(\hbar\omega')^2}}}|\}
\end{eqnarray}
By defining $y^2=u^2+a^2$, $I_2$ and $I_3$ are obtained as
\begin{eqnarray}
I_2&=&\frac{1}{4\hbar\omega'}\ln|\frac{u-\frac{\hbar\omega'}{2}}{u+\frac{\hbar\omega'}{2}}|\nonumber\\
I_3&=&\frac{1}{(\hbar\omega')^2u}+\frac{1}{(\hbar\omega')^3}\ln|\frac{u-\frac{\hbar\omega'}{2}}{u+\frac{\hbar\omega'}{2}}|
\end{eqnarray}
Using the above expressions for $I_1, I_2, I_3$, and $I_4$, it is easy to
prove Eqs.~(\ref{GG}) and (\ref{HH}).


\begin{thebibliography}{99}

\bibitem{Xu}
M. Xu, T. Liang, M. Shi, and H. Chen, Chemical Reviwes, {\bf 113}, 3766 (2013).

\bibitem{Geim}
A. K. Geim, and I. V. Grigorrieva, Nature (London) {\bf 499}, 419 (2013).

\bibitem{wang12}
Q. H. Wang, K. Kalantar-Zadeh, A. Kis, J. N. Coleman, and M. S. Strano, Nat. Nanotechnol. {\bf 7}, 699 (2012).

\bibitem{Neto09}
A. H. Castro Neto, F. Guinea, N. M. R. Peres, K. S. Novoselov and
A. K. Geim, Rev. Mod. Phys. \textbf{81}, 109 (2009).

\bibitem{Hasan10}
M. Z. Hasan, C. L. Kane, Rev. Mod. Phys. \textbf{82}, 3045 (2010).

\bibitem{Tibook}
Shun-Qing Shen,\emph{Topological insulator: Dirac equation in
condesed matters}, Springer (2012).

\bibitem{Zhang13}
T. Zhang, J. Ha, N. Levy, Y. Kuk, and J. Stroscio, Phys. Rev. Lett. \textbf{111}, 056803 (2013).

\bibitem{Lu10}
H.-Z. Lu, W.-Y. Shan, W. Yao, Q. Niu, and S.-Q. Shen, Phys. Rev. B \textbf{81}, 115407(2010).

\bibitem{Li10}
H. Li, L. Sheng, D. N. Sheng, and D. Y. Xing, Phys. Rev. B {\bf
82}, 165104 (2010).

\bibitem{HLi12}
H. Li, L. Sheng, and D. Y. Xing, Phys. Rev. B \textbf{85}, 045118(2012).

\bibitem{Mak10}
 K. F. Mak, C. Lee, J. Hone, J. Shan, and T. F.Heinz, Phys. Rev. Lett. \textbf{105}, 136805 (2010).

\bibitem{mak12}
K. F. Mak, K. He, J. Shan, and T. F. Heinz, Nat. Nanotechnol. {\bf 7}, 494 (2012).

\bibitem{mak13}
K. F. Mak, K. He, C. Lee, G. H. Lee, J. Hone, T. F. Heinz, and J. Shan, Nat. Mat. {\bf 12}, 207 (2013).

\bibitem{cui12}
H. Zeng, J. Dai, W. Yao, D. Xiao, and X. Cui, Nat. Nanotechnol.  \textbf{7}, 490 (2012).

\bibitem{cao12}
T. Cao, G. Wang, W. Han, H. Ye, C. Zhu, J. Shi, Q. Niu, P. Tan, E. Wang, B. Liu, and J. Feng, Nature Commun. \textbf{3}, 887 (2012).

\bibitem{wu13}
S. Wu, J. S. Ross, G. B. Liu, G. Aivazian, A. Jones, Z. Fei, W. Zhu, D. Xiao, W. Yao, D. Cobden, and X. Xu, Nat. Phys. {\bf 9}, 149 (2013).

\bibitem{Rycerz07}
A.Rycerz, J.Tworzydlo, and C. W. J. Beenakker, Nat. Phys. \textbf{3}, 172 (2007).

\bibitem{xiao07}
D.Xiao, W.Yao, and Q. Niu,  Phys. Rev. Lett. \textbf{99}, 236809 (2007).

\bibitem{yao08}
W.Yao, D. Xiao, and Q. Niu, Phys. Rev. B \textbf{77}, 235406 (2008).

\bibitem{xiao12}
Di Xiao, Gui-Bin Liu, W. Feng, X. Xu, and W. Yao, Phys. Rev. Lett. \textbf{108}, 196802 (2012).

\bibitem{Rostami13}
H. Rostami, A. G. Moghaddam, and R. Asgari, Phys. Rev. B
\textbf{88}, 085440 (2013).

\bibitem{Liu13}
G. -B. Liu, W. -Y. Shan, Y. Yao, W. Yao, D. Xiao, Phys. Rev. B \textbf{88}, 085433 (2013).

\bibitem{Kormanyos13}
A. Kormanyos, V. Zolyomi, N. D. Drummond, P. Rakyta, G. Burkard,
and V. I. Fal'ko, Phys. Rev. B \textbf{88}, 045416 (2013).

\bibitem{Carvalho13}
A. Carvalho, R. M. Ribeiro, and A. H. Castro Neto, Phys. Rev. B
\textbf{88}, 115205 (2013).

\bibitem{Li12}
Zhou Li and J. P. Carbotte, Phys. Rev. B \textbf{86}, 205425(2012).



\bibitem{Shan10}
W.-Y. Shan, H.-Z. Lu, and S.-Q. Shen, New J. Phys. \textbf{12},
043048 (2010).

\bibitem{Kim12}
M. Kim, C. H. Kim, H.-S. Kim, and J. Ihm, PNAS {\bf 109}, 671
(2012).

\bibitem{peres13}
N. M. R. Peres, and J. E. Santos, J. Phys.: Condens. Matter {\bf
25} 305801 (2013).

\bibitem{Lu13}
Hai-Zhou Lu, An Zhao,and Shun-Qing Shen, Phys. Rev. Lett.
\textbf{111},146802 (2013).

\bibitem{Stauber08}
T. Stauber, N. M. R. Peres, and A. K. Geim, Phys. Rev. B
\textbf{78} 085432 (2008).

\bibitem{Tse11}
Wang-Kong Tse and A. H. MacDonald, Phys. Rev. B \textbf{84},205327(2011).

\bibitem{Louie06}
Steven G. Louie and Marvin L. Cohen,\emph{Conceptual Foundations of Materials:
A Standard Model for Ground- and Excited-State Properties}, Elsevier (2006).

\bibitem{Gomez13}
A. C.-Gomez, R. Rold\'{a}n, E. Cappelluti, M. Buscema, F. Guinea, H. S. J. van der Zant, and G. A. Steele, Nano Lett., {\bf 13}, 5361 (2013).

\bibitem{yao09}
Wang Yao, Shengyuan A. Yang, and Qian Niu, Phys. Rev. Lett.
\textbf{102}, 096801 (2009).

\bibitem{Zigler07}
K.Zigler, Phys. Rev. B \textbf{75}, 233407(2007).

\bibitem{Ferreira11}
A. Ferreira, J. V.-Gomes, Y. V. Bludov, V. Pereira, N. M. R.
Peres, and A. H. Castro Neto ,Phys. Rev. B \textbf{84}, 235410
(2011).

\bibitem{zhang09}
H. J. Zhang, C. X. Liu, X. L. Qi, X. Dai, Z. Fang, and S. C. Zhang, Nat. Phys. \textbf{5}, 438 (2009).

\bibitem{jahn}
kh. Jahanbani and R. Asgari, Eur. Phys. J. B {\bf 73}, 247 (2010).
\end{thebibliography}
\end{document}